\documentclass[aps,reprint,amsmath, amssymb,groupedaddress,floatfix]{revtex4-2}
\pdfoutput=1
\usepackage{natbib}
\usepackage{amsbsy}
\usepackage{graphicx}
\usepackage{amsmath}
\usepackage{bm}
\usepackage{color,soul}
\usepackage{hyperref}
\usepackage{float}
\usepackage{csquotes}
\usepackage{xcolor}
\usepackage{color,soul}
\hypersetup{colorlinks=true, urlcolor=blue, citecolor=blue}
\usepackage{url}

\newcommand{\nbh}{2H-NbSe$_2$}
\newcommand{\nb}{NbSe$_2$}
\newcommand{\wtt}{1T$^\prime$-WTe$_2$}
\newcommand{\wt}{WTe$_2$}

\DeclareUnicodeCharacter{2212}{-}

\begin{document}
\preprint{APS}
\title{Role of interface hybridization on induced superconductivity in \wtt~and \nbh~heterostructures}
 \author{Anirban Das$^{1,2}$, Bent Weber$^{3,4}$, Shantanu Mukherjee$^{1,2,5}$}
 \affiliation{$^1$Department Of Physics, Indian Institute Of Technology Madras, Chennai, 600036, India\\
 $^2$Computational Materials Science Group IIT Madras, Chennai, Tamil Nadu 600036, India\\
 $^3$Division of Physics and Applied Physics, School of Physical and Mathematical
Sciences, Nanyang Technological University, Singapore 637371, Singapore\\
$^4$ARC Centre of Excellence for Future Low-Energy Electronics Technologies (FLEET), School of Physics, Monash University, Clayton VIC 3800, Australia\\
$^5$Quantum Centres in Diamond and Emergent Materials (QCenDiem)-Group IIT Madras Chennai, Tamil Nadu 600036, India}
 \begin{abstract}
Heterostructures between two dimensional quantum spin Hall insulators (QSHI) and superconducting materials can allow for the presence of Majorana Fermions at their conducting edge states. Although a strong interface hybridization helps induce a reasonable superconducting gap on the topological material, the hybridization can modify the material's electronic structure. In this work, we utilize a realistic low energy model with tunable interlayer hybridization to study the edge state physics in a heterostructure between monolayer quantum spin Hall insulator \wtt~and s-wave superconductor \nbh. We find that even in the presence of strong inter-layer hybridization that renders the surface to become conducting, the edge state shows a significantly enhanced local density of states and induced superconductivity compared to the surface. We provide an alternate heterostructure geometry that can utilize the strong inter-layer hybridization and realize a spatial interface between a regime with a clean QSHI gap and a topological conducting edge state.     
\end{abstract}
\maketitle

\section{Introduction}
Among all the polytropes of monolayer \wt, the 1T$^\prime$ structure is being extensively studied because of its non-trivial topological properties \cite{heising1999structure,eda2012coherent,qian2014quantum,ugeda2018observation,chen2018large,jia2022tuning}. Band structure calculations have revealed that it is a quantum spin Hall insulator (QSHI), where counter-propagating spin momentum locked electrons move along the edge, forming helical edge states  \cite{tang2017quantum,fei2017edge,qian2014quantum}. Recent studies have found that these materials could be an excellent platform to generate quasi-particle bound states like Majorana fermions at the edges of the monolayer via proximity-induced superconductivity with a conventional superconductor \cite{fu2008superconducting}. It has already been established that the heterostructure of \wt~and \nb~leads to an induced superconducting order parameter in \wt, that is quite robust and the induced superconducting gap is enhanced at the topological edge states of the \wt~monolayer \cite{tao22,li2018proximity,lupke2020proximity,huang2018inducing}. The strength of the induced superconducting pairing at the edges depends intricately on the electronic structure of the edge state and the interlayer coupling between the materials.

When a monolayer \wtt~is grown on \nbh~substrate, the van der Waals interaction between \wt~and \nb~determines the strength of interface coupling. It is important to explore the role of interface coupling on the underlying electronic structure and induced superconducting gap of the monolayer topological insulator, since these material parameters lay the foundation for possible nucleation of Majorana bound states at the edges of the quantum spin Hall insulator \cite{fu2008superconducting, PhysRevLett.104.056402}. The role of the interface hybridization attains further significance in engineered heterostructures where the hybridization is tuned by insertion of decoupled layers between the hybridizing materials\cite{fang2014strong,lee2019planar,bonell2020control,binder2019upconverted}.

\begin{figure}[t]
    \centering
    \includegraphics[width =\columnwidth]{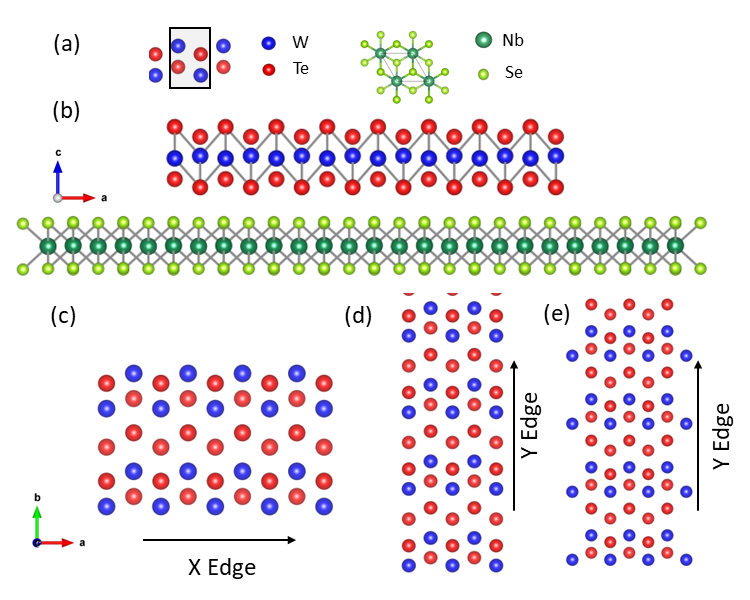}
    \caption{(a) The unit cells of \wtt~and \nbh. (b) Schematics of the heterostructure formation showing the monolayer of \wtt~over \nbh. The alignments of W and Te ions at the three different edges of \wtt~are shown in (c), (d), and (e).}
    \label{mdoel}
\end{figure}

In this work, we present a theoretical model of \wtt/\nbh~heterostructure in order to explore the influence of interface coupling strength on the induced superconducting gap and topological nature of the edge state in \wtt. We use a multi-orbital real-space tight-binding Hamiltonian relevant to the low energy properties of \wt~and \nb~to model the heterostructure and study the evolution of the normal state electronic structure and induced superconductivity as a function of the interface coupling strength. 

The study reveals that the low-energy electronic structure is highly susceptible to weak interface hybridization between the \wtt~and \nbh~electronic states. This leads to the smearing out of the QSHI gap in the bulk and the formation of hybridized conducting edge states. Considering three examples of edge terminations in \wt, we show that the nature of the edge state is dependent on both the termination direction and the terminating \wtt~ions. 

Our real space self-consistent calculations of the superconducting order indicate that the induced spin singlet state can be present both at the surface and edge of \wtt~in the presence of a finite interface coupling. We discuss how the induced order varies as a function of interface coupling and edge termination that are naturally present in experimental conditions. 

We also evaluate the induced effective spin-triplet superconducting order parameter proposed to be present in the presence of linearly dispersing Dirac bands constituting the edge states of quantum spin  Hall insulators \cite{fu2008superconducting}. We find that this chiral triplet order can survive at \wtt~edges even in the presence of interface hybridization which is indicative of the presence of linearly dispersing bands coinciding with the bulk states in the hybridized heterostructure at the Fermi level. 

In the final section, we study the effect of inducing a superconducting gap on \wtt~in a heterostructure where the superconductor has dimensions smaller than the monolayer QSHI material. We show that although the strong inter-layer hybridization can lead to the monolayer QSHI in contact with the superconductor to become metallic and have modified topological properties, this geometry can nevertheless lead to the formation of a clean edge separating regions with different topological properties.  

\begin{figure}[t]
    \centering
    \includegraphics[width=\columnwidth]{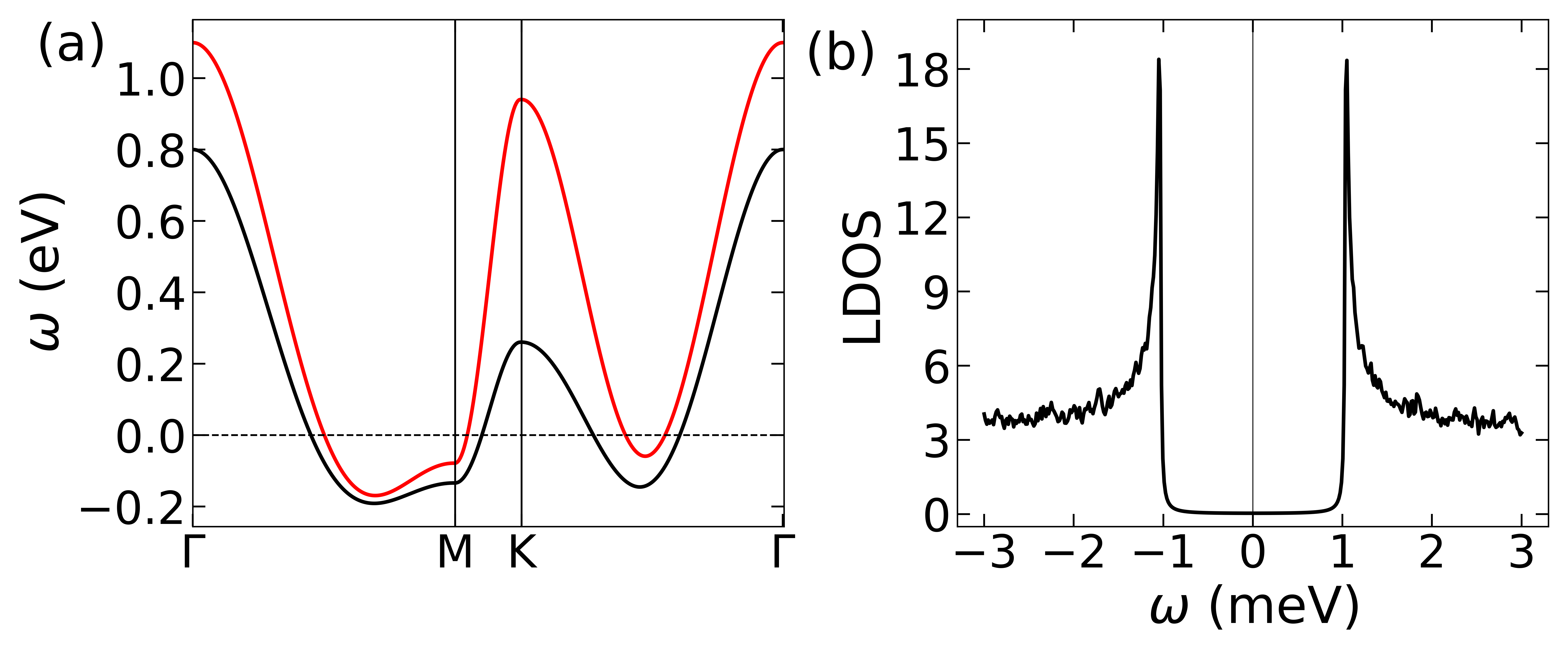}
    \caption{(a) Low energy electronic bands of \nbh~over the high symmetry path showing the two bands formed from bilayer interactions. (b) Calculated local density of states (LDOS) in the superconducting state showing the gap size of $\sim$ 1.1 meV for pristine \nbh.}
    \label{nbse2}
\end{figure}

\section{model}

 The \wtt/\nbh~heterostructure has been modeled using a real space mean field Hamiltonian\cite{githublink}.
 Fig. \ref{mdoel}(a) displays the unit cell structures of \wtt~and \nbh. Fig. \ref{mdoel}(b) depicts a schematic of the heterostructure, where blue and red spheres represent W and Te ions, and deep green and light green spheres represent Nb and Se ions, respectively. Finally, Fig. \ref{mdoel}(c), (d), and (e) show different edges of the \wtt~structure. The relevant problem for the superconducting state has been solved using a self-consistent Bogoliubov-de Gennes (BdG) formalism \cite{de1966superconductivity}(See appendix A). The Hamiltonian contains four separate parts given by, 
	
\begin{eqnarray}
	\mathcal{H}&=&\mathcal{H}^{0}_{N}+\mathcal{H}^{0}_{W}+\mathcal{H}_{W-N} +\mathcal{H}^{N}_{SC}
\end{eqnarray} 

The individual terms of the Hamiltonian are,
	
\begin{eqnarray}
	\mathcal{H}^{0}_{N}&=&\sum_{iljl'\sigma}t^{ll'}_{ij}c^{\dag}_{il\sigma}c_{jl'\sigma} + h.c \nonumber\\
	\mathcal{H}_{W-N}&=&t_{\perp} \sum_{ij\sigma} c^{\dag}_{i2\sigma}d_{j1\sigma} + h.c \nonumber\\
	\mathcal{H}^{0}_W&=&\sum_{\mu\nu ij\sigma\sigma'}t^{ij}_{\mu\nu\sigma\sigma'}d^{\dag}_{i\mu\sigma}d_{j\nu \sigma'} + h.c \nonumber\\
	\mathcal{H}^{N}_{SC}&=&  \sum_{il}\Delta^{N}_{il} c^{\dag}_{il\uparrow}c^{\dag}_{il\downarrow}+ h.c\nonumber
\end{eqnarray}

\begin{figure}[t]
    \centering
    \includegraphics[width=0.32\columnwidth]{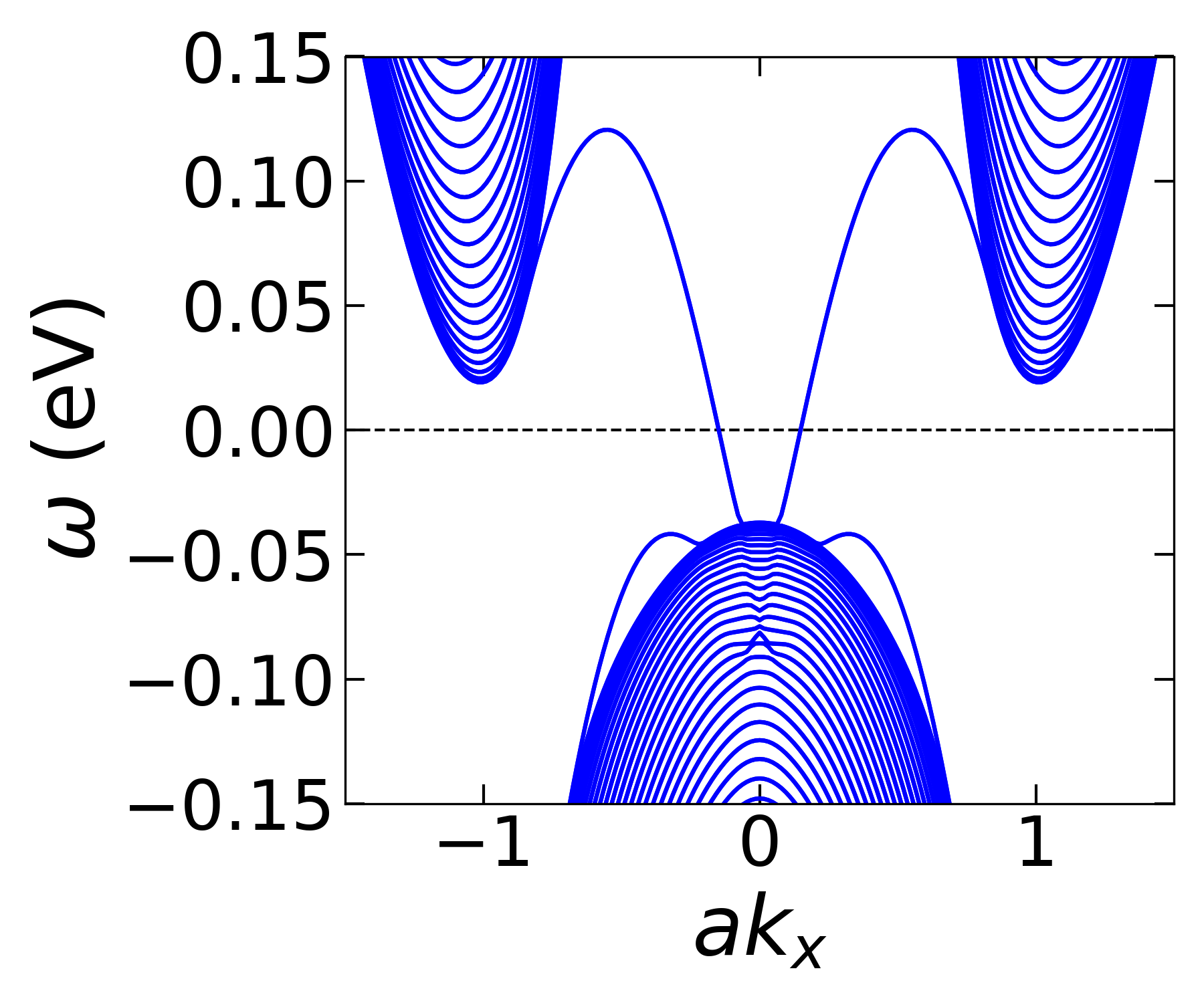}
    \includegraphics[width=0.32\columnwidth]{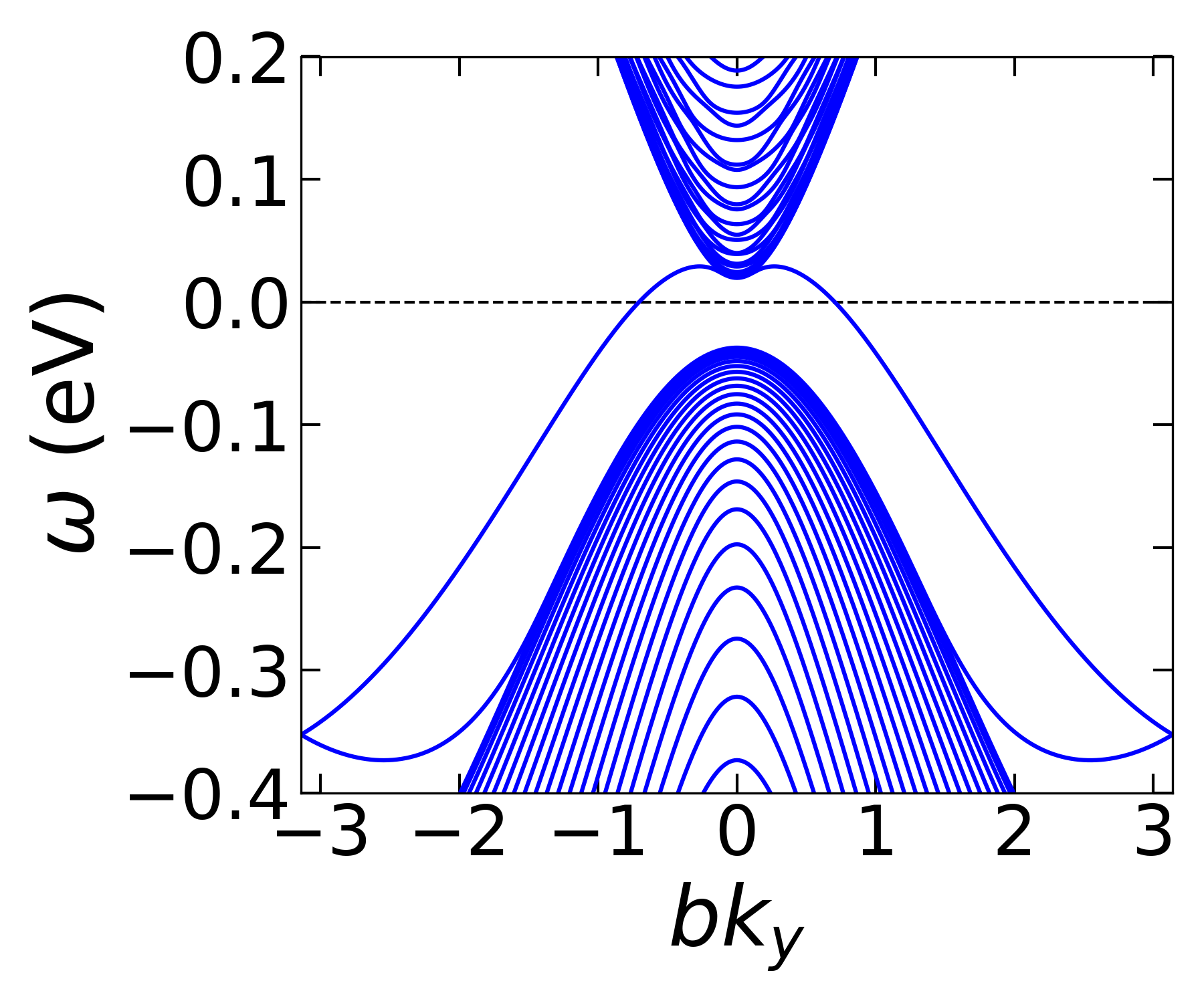}
    \includegraphics[width=0.32\columnwidth]{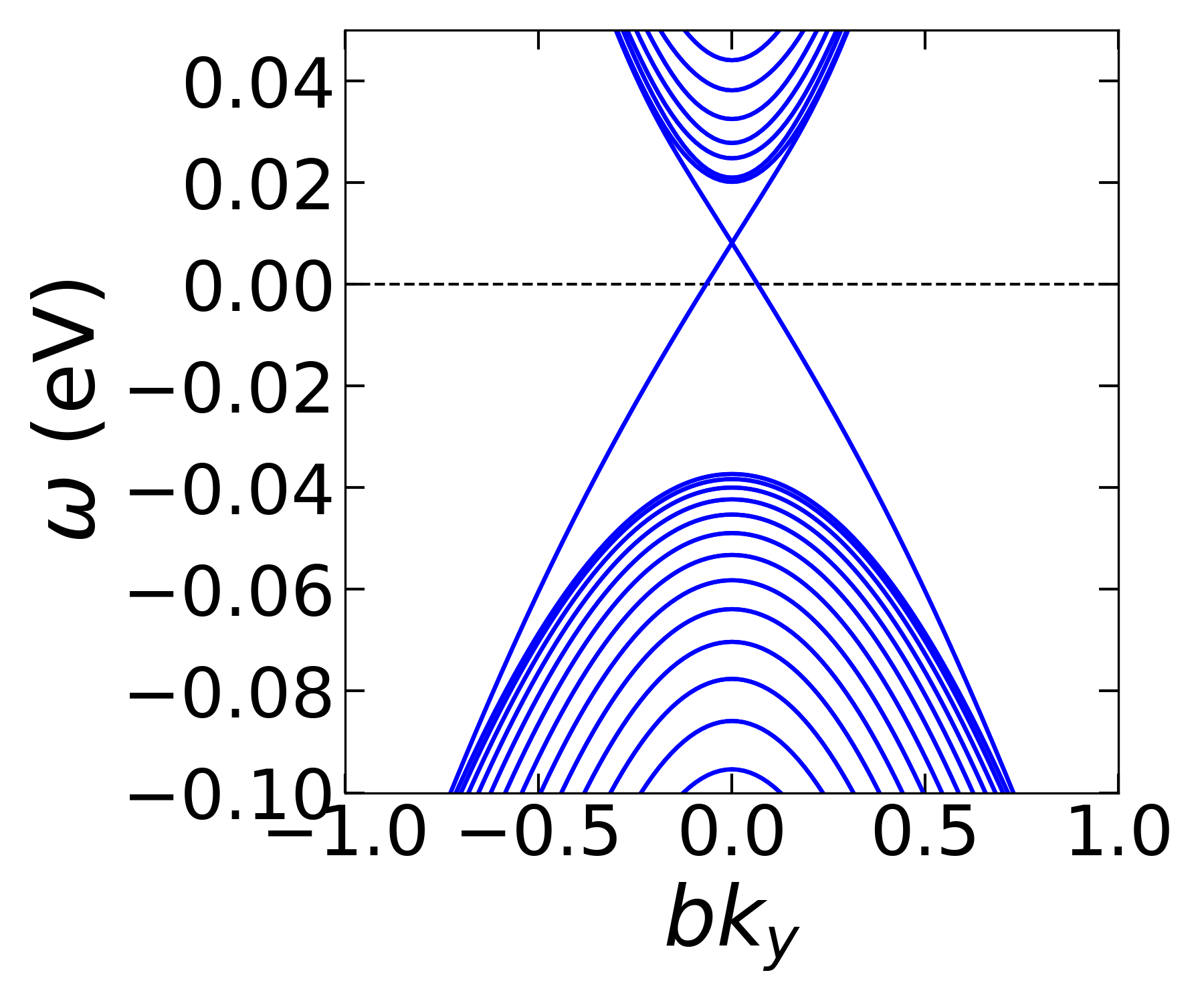}\\
    \includegraphics[width=0.32\columnwidth]{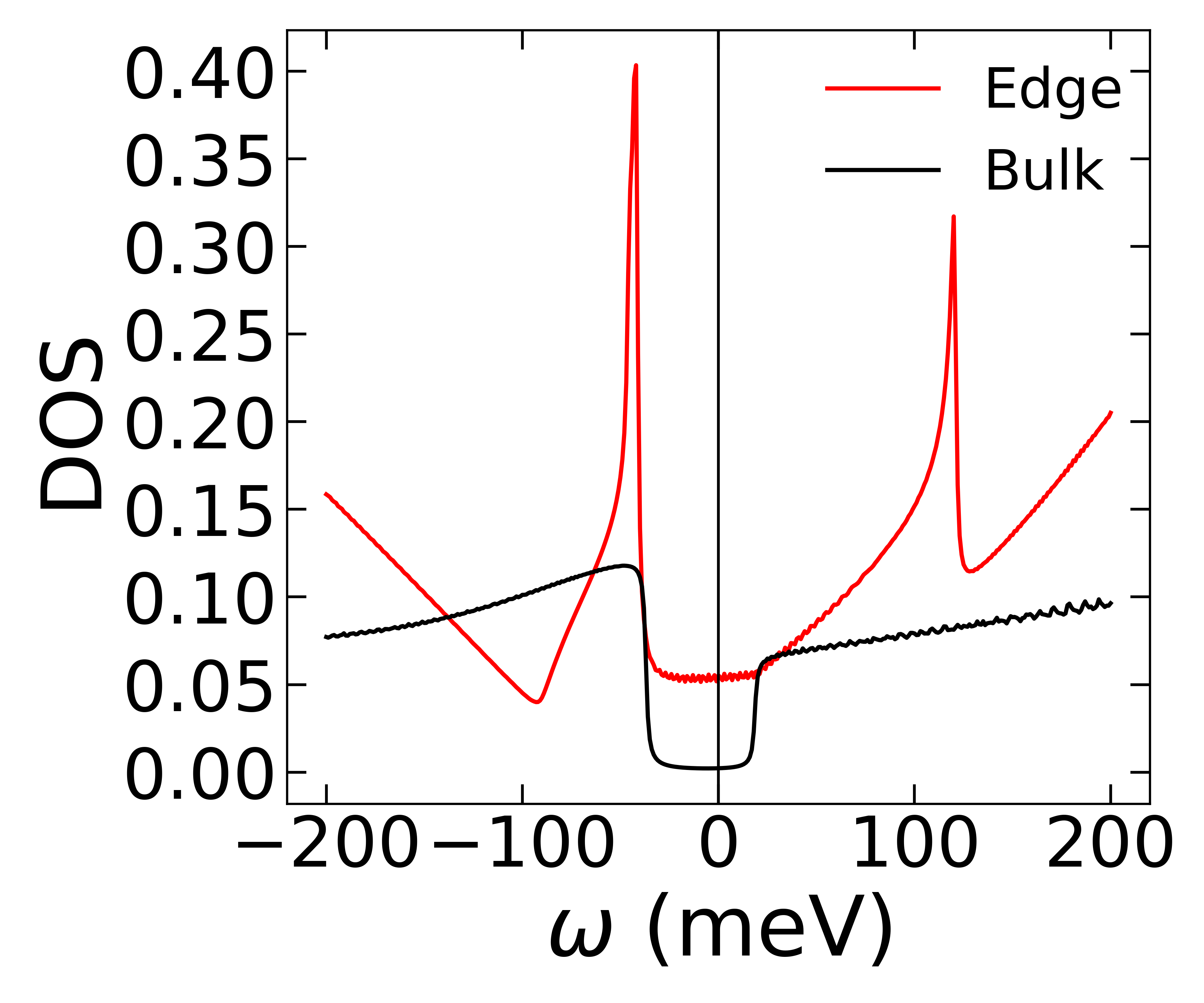}
    \includegraphics[width=0.32\columnwidth]{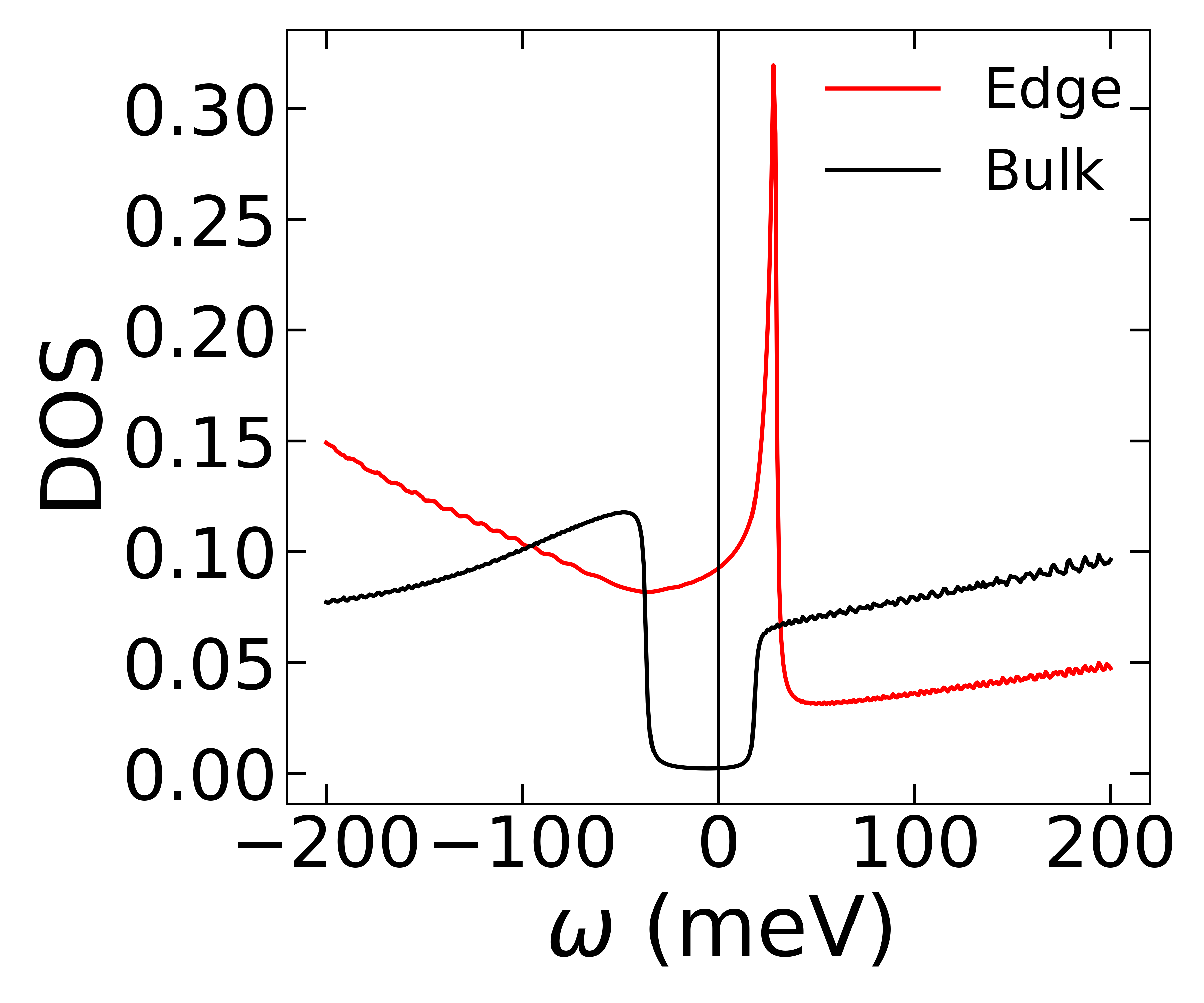}
    \includegraphics[width=0.32\columnwidth]{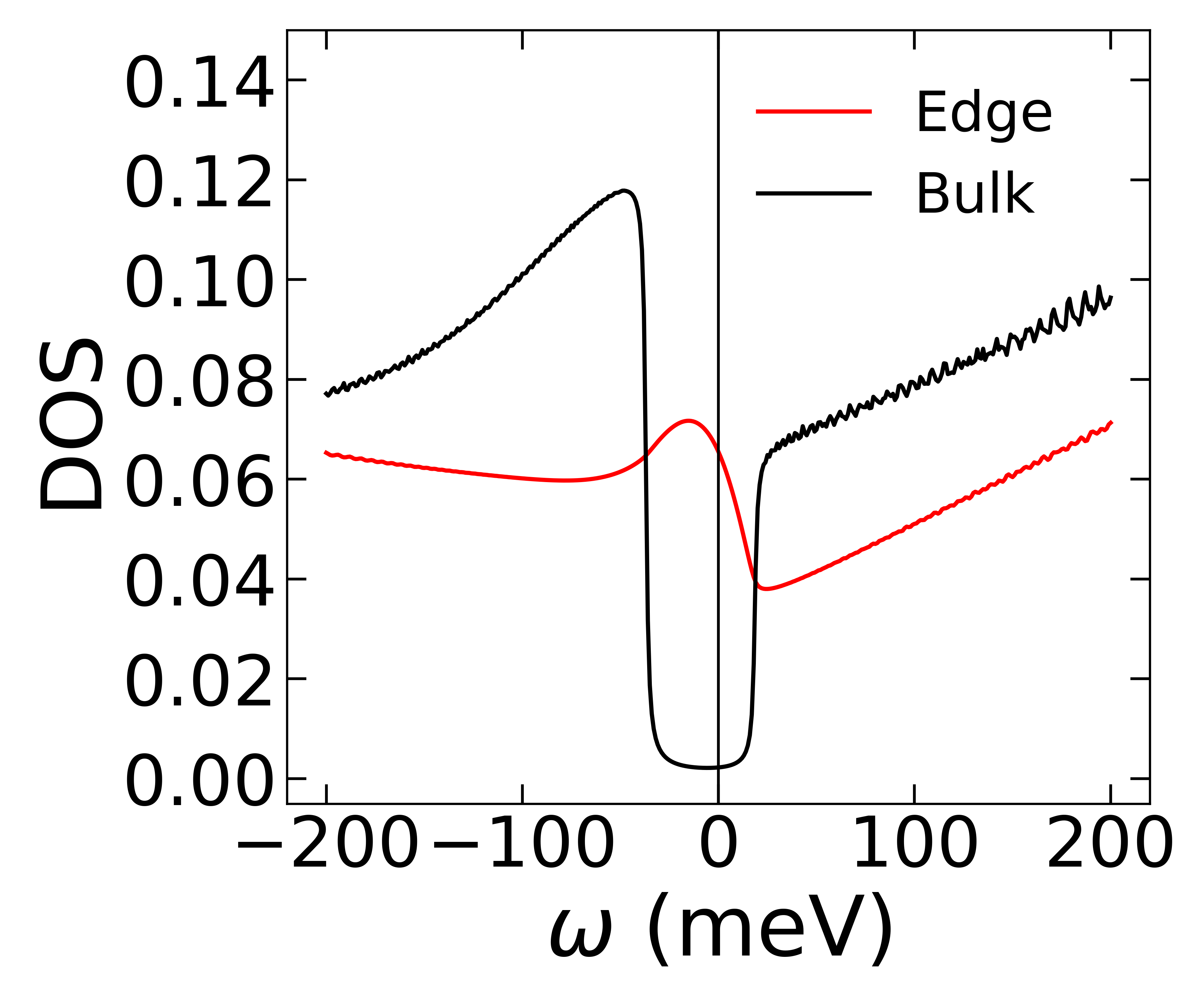}
    \caption{Band diagrams of pristine \wtt~for three different terminations showing the edge state bands crossing the Fermi energy ($E_F$), indicated by the horizontal black line (top panel). Calculated normal state DOS are shown in the lower panel. The red line indicates the DOS of the conducting edge whereas the black line is showing the bulk DOS indicating the quantum spin Hall gap of size $\sim$ 56 meV. }
    \label{pristine_bands}
\end{figure}

In the above $\mathcal{H}^{N}_{0}$ is the non interacting Hamiltonian for \nbh~with the operators $(c^{\dag}_{il\sigma},c_{jl\sigma})$ representing the creation and annihilation operators respectively at the site $i$ and $j$ (considering a single d$_{3z^2}$ orbital at each site), $l=(1,2)$ is the layer index of this bilayer material, and $\sigma=(\uparrow,\downarrow)$ is the spin index. This tight binding Hamiltonian is generated by a basis transformation to orbital and layer basis, from a two band Hamiltonian that has been studied previously to explain ARPES \cite{rahn2012gaps} and STM experiments  \cite{gao2018atomic,soumyanarayanan2013quantum} on 2H-NbSe$_2$. The normal state low energy electronic for \nbh~is shown in Fig.~\ref{nbse2}(a). The electronic structure agrees well with the earlier tight binding \cite{doran1978tight} and DFT calculations \cite{johannes2006fermi} and ARPES experiments \cite{rahn2012gaps}.

\wtt~is a Quantum spin Hall insulator (QSHI) having spin-momentum locked helical edge states and a spin hall gap of $\sim$ 56 meV at the bulk. It has 6 atoms in a rectangular unit cell, however as discussed in \cite{lau19, Choe16, Muechler16, Hu21} the low energy bands are dominated by two Te $p_x$ orbital and two W $d_{x^2 - y^2}$ and $d_{z^2}$ orbitals because of the distortion in the unit cell. We model the electronic structure of \wt~using a real space version of 8 orbital Hamiltonian where the 4 atom basis is doubled in the presence of a Rashba spin orbital interaction. In $\mathcal{H}^{0}_W$ the indices $i, j$ refer to the unit cell, and $\mu,\nu=(1,2,3,4)$ represent the 2 W and 2 Te ions present within each unit cell respectively. This momentum space version of the model was derived in \cite{lau19} and is generated by a combination of DFT and fitting to ARPES experiments. The model provides reasonable agreement with the low energy electronic structure \cite{Choe16, PhysRevX.6.041069} including the QSHI observed in ARPES \cite{tang2017quantum} and STM experiments \cite{tang2017quantum,maximenko2022nanoscale}.

\begin{figure}[t]
    \centering
    \includegraphics[width=\columnwidth]{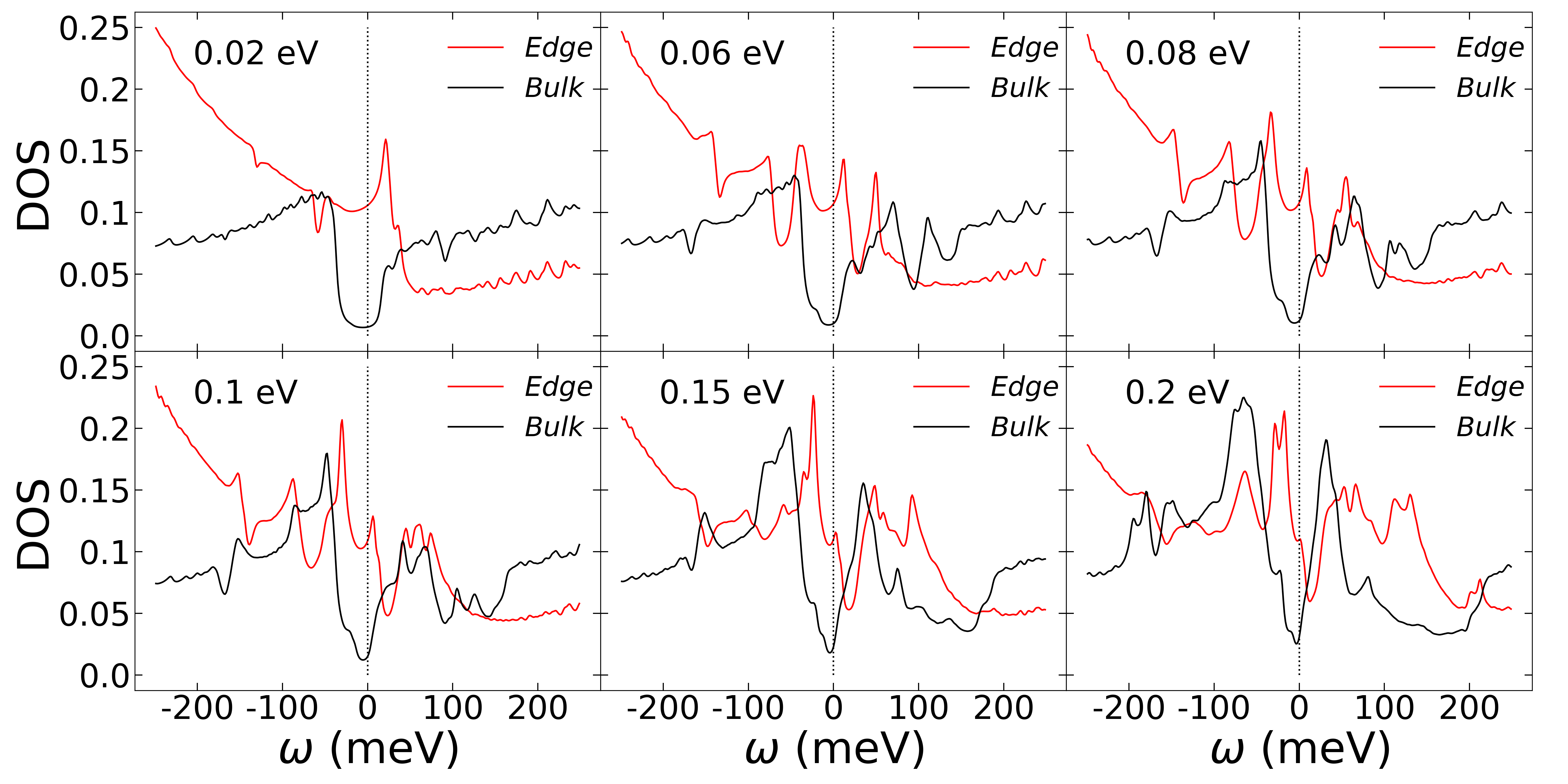}
    \caption{\textbf{Normal state DOS of Y-edge:} Normal state density of states (DOS) of \wtt~for different interface coupling strengths (t$_\perp$) ranging from 0.02 eV to 0.2 eV as labeled in the plots. The red line represents the DOS for the edge unit cell whereas the black line corresponds to the bulk. The dotted vertical line in each plot indicates the Fermi energy(E$_F$).}
    \label{normal}
\end{figure}

The edge was modeled in \cite{lau19} using an open boundary condition for various terminations and directions perpendicular and parallel to atom chain directions that are naturally present in the structure of monoclinic \wtt. The edge state electronic structure is reproduced in Fig \ref{pristine_bands} and shows the edge bands of \wtt~for three different edge terminations \cite{lau19}. The edge state electronic structure in Fig. \ref{pristine_bands} (a) is for an X-edge with W termination, Fig. \ref{pristine_bands} (b) is for  W terminated smooth Y-edge, and Fig. \ref{pristine_bands} (c) shows the electronic structure at a W terminated saw-tooth type Y-edge \cite{lau19}. The corresponding schematic diagram of these edges has been shown in Fig. \ref{mdoel} (c), Fig. \ref{mdoel} (d), and Fig. \ref{mdoel} (e) respectively. Note that among the three cases discussed in this work, a Dirac like spectrum is obtained only for the W terminated saw-tooth type Y-edge. The local density of states (LDOS) for the three edges that we hereafter refer to as edge (a), edge (b), and edge (c) are shown in Figs. \ref{pristine_bands} (d), \ref{pristine_bands} (e), and \ref{pristine_bands} (f) respectively. We find that edge (a) leads to a constant density of states at low energies indicative of a linear dispersion. Note that the density of states is anisotropic between the various edge terminations that can be expected to have implications for both the magnitude and structure of any induced superconducting gap.

Our real space model transforms the \wtt~into a lattice model with open boundary conditions applied to various edges and termination directions. Although this is a simplified model for the edge as in real systems we are likely to have edge regions that are more disordered, this model has been reasonably successful in reproducing the essential features observed in recent STM experiments \cite{tao22}.     

\nbh~is a conventional superconductor with a critical temperature $\sim 7.2K$ \cite{wilson1975charge}. Fig. \ref{nbse2} (a) shows the band structure of \nbh. We calculate the superconducting order parameter $\Delta^{N}_{il}=V\langle c_{il\uparrow}c_{il\downarrow}\rangle $ self consistently at each lattice site by solving the BdG equations. The on-site s-wave pairing strength $V$ generates a superconducting gap of $\Delta \sim 1.1 $ meV in agreement with the experimental results on pristine 2H-NbSe$_2$ (see Fig. \ref{nbse2} b) \cite{Guillamprl08,klemm2015pristine}. Note that the model ignores the additional contribution from a 3Q charge density wave order since in the presence of disorder and heterostructure, local probe experiments find that CDW order to be strongly suppressed \cite{tao22}.

\begin{figure}[t]
    \centering
    \includegraphics[width=\columnwidth]{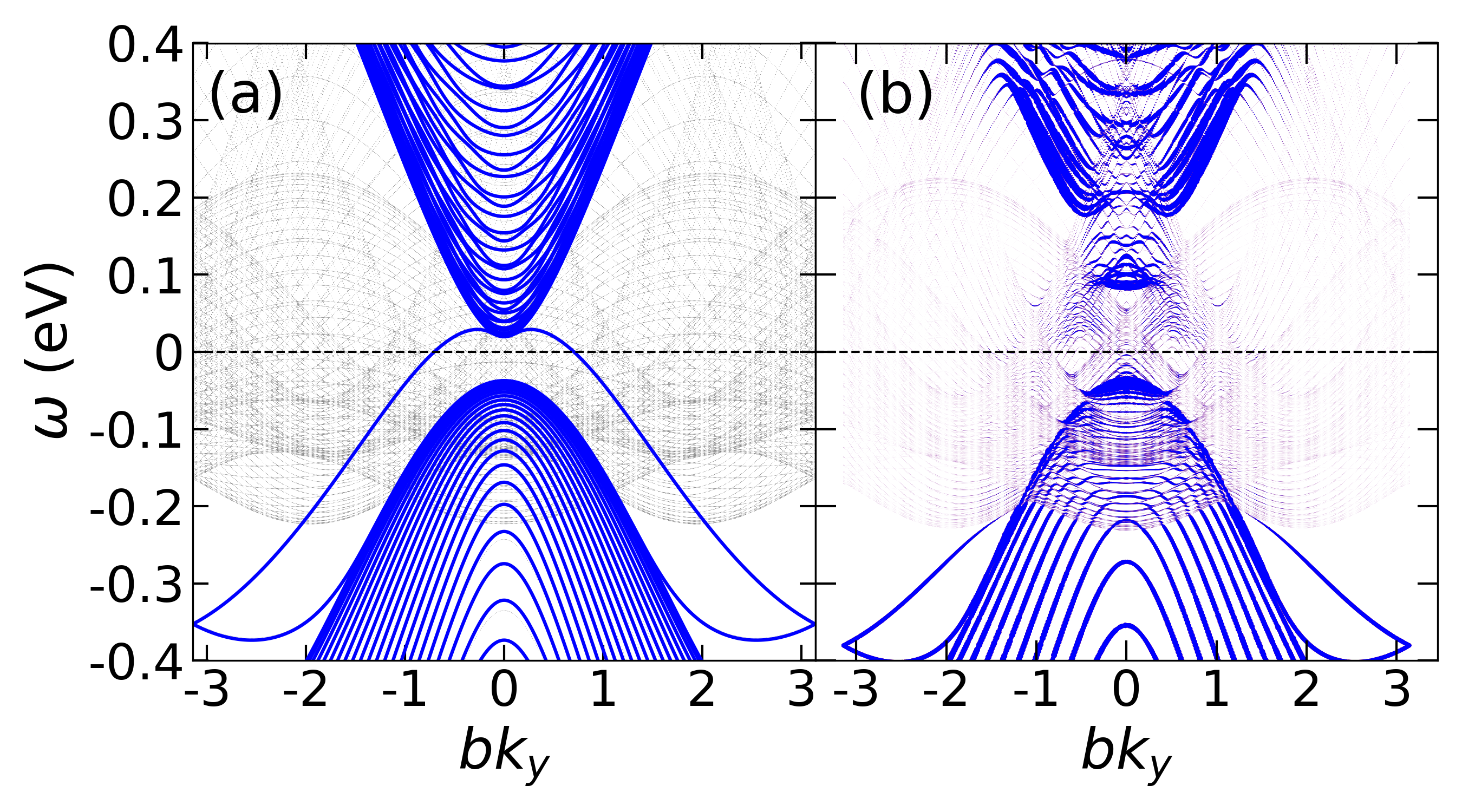}
    \caption{(a) Calculated Band structure of pristine \wtt~(blue) showing the surface band crossing the Fermi Energy. The grey lines in the background showing the pristine \nbh bandstructure within the folded Brillouin zone.} (b) Hybridized band structure of \wtt~with interface coupling 0.15 eV.
    \label{bulk_bands}
\end{figure}

 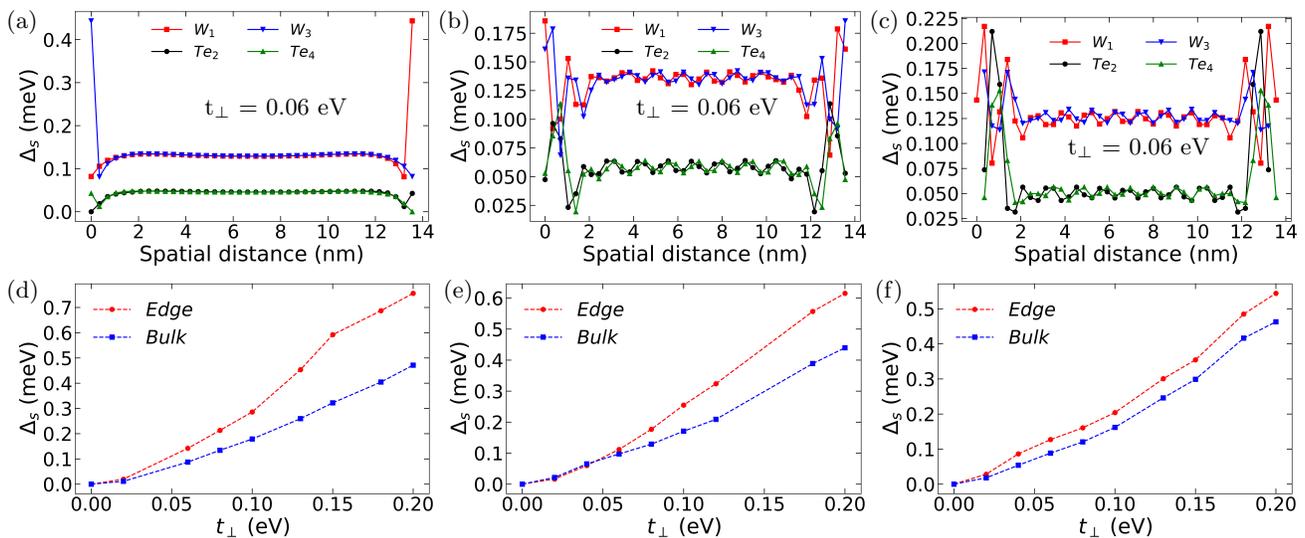
\begin{figure*}[t]
    \centering
    \includegraphics[width=0.32\textwidth]{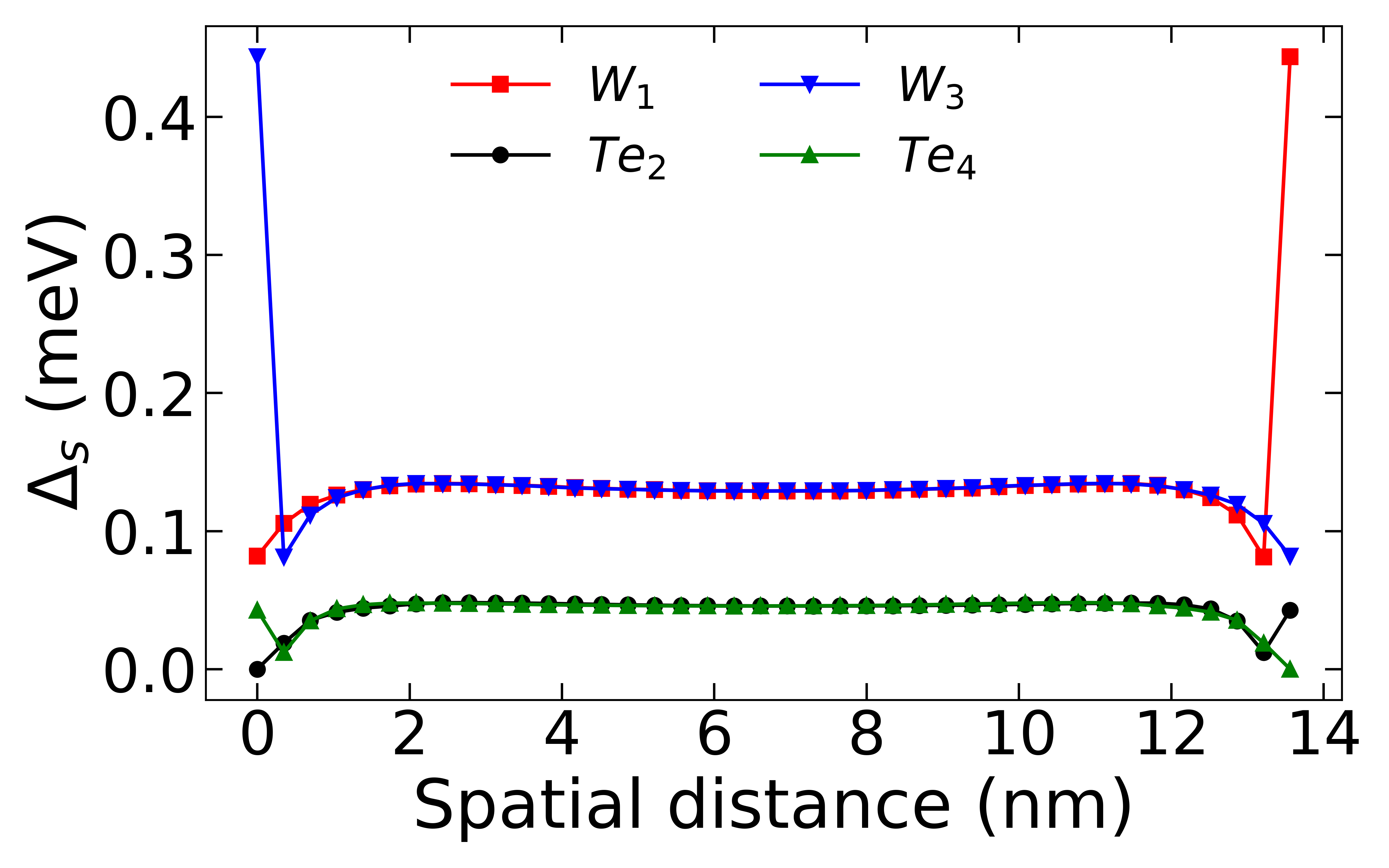}
    \put(-165,93){(a)}
    \put(-90,60){t$_\perp$ = 0.06 eV}
    \includegraphics[width=0.32\textwidth]{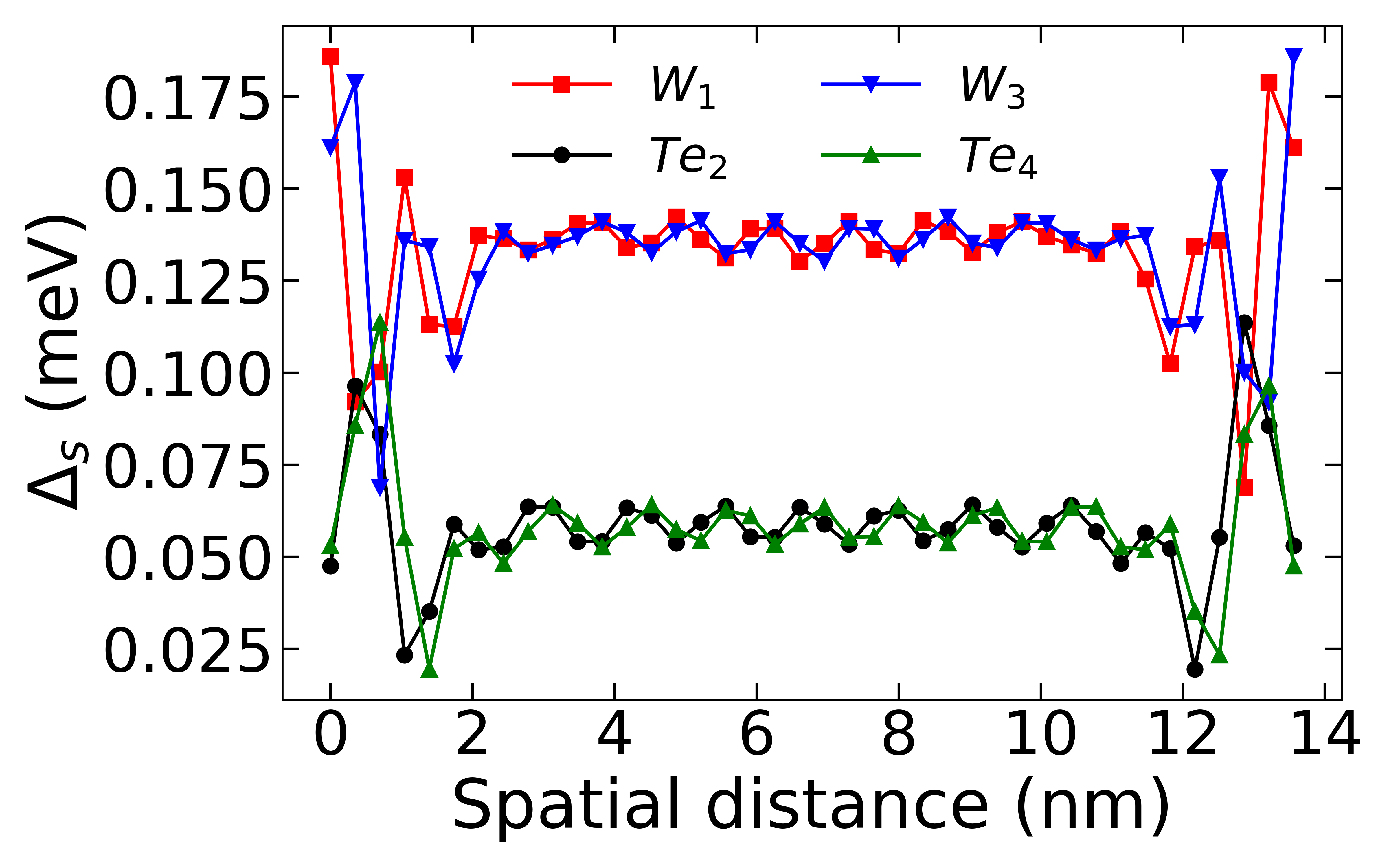}
    \put(-165,93){(b)}
    \put(-90,60){t$_\perp$ = 0.06 eV}
    \includegraphics[width=0.32\textwidth]{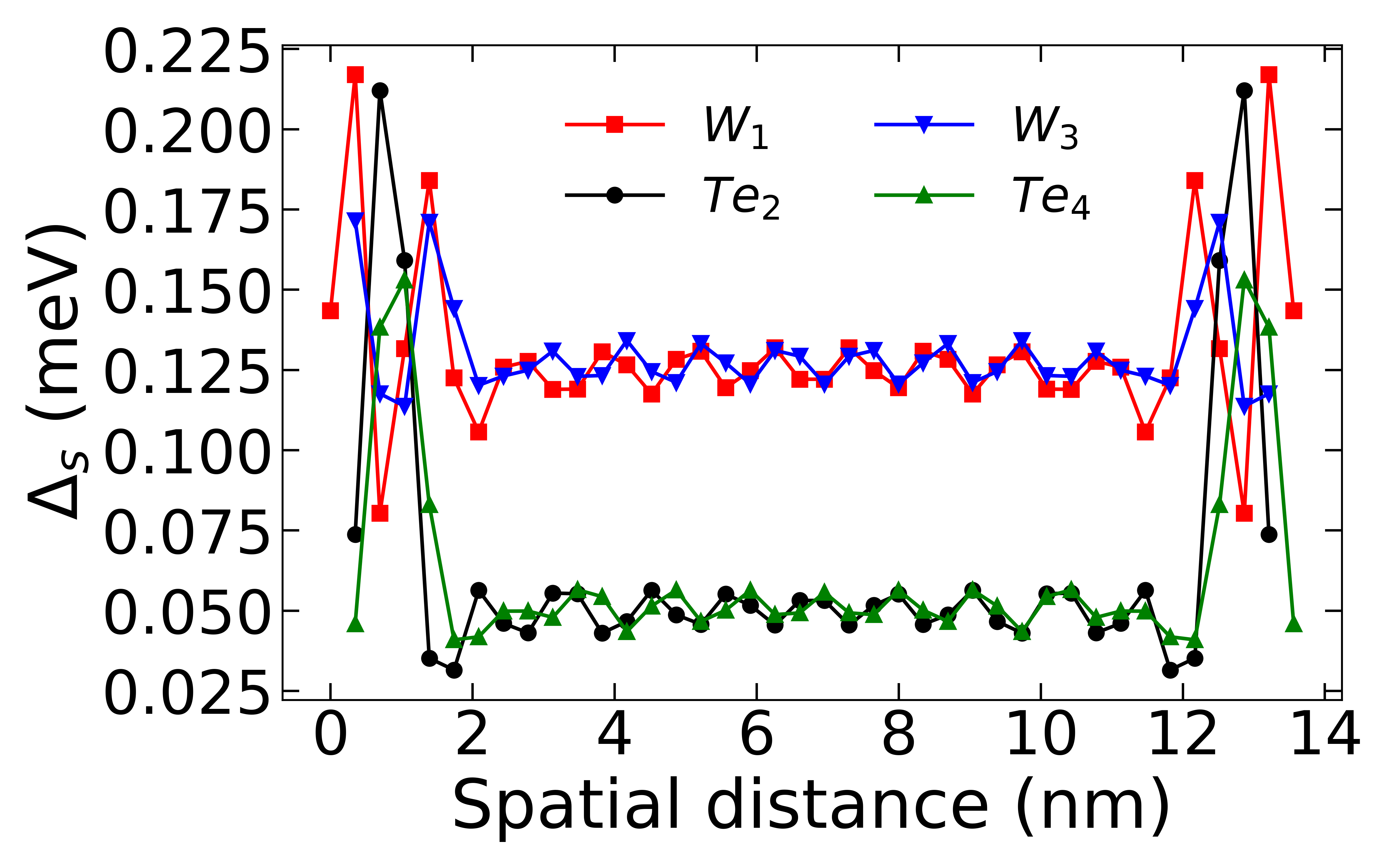}
    \put(-165,93){(c)}
    \put(-90,45){t$_\perp$ = 0.06 eV}\\
    \includegraphics[width=0.32\textwidth]{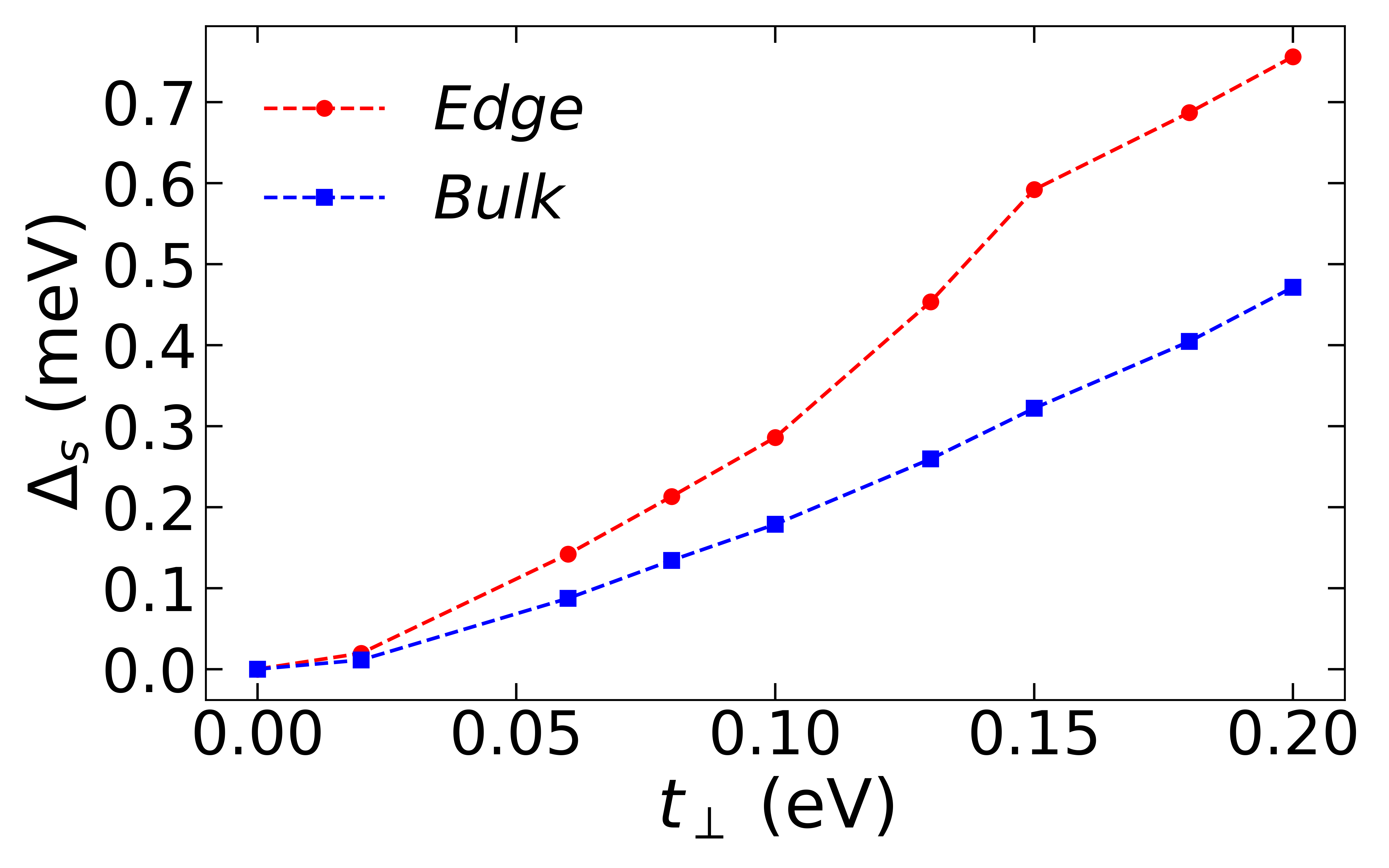}
    \put(-165,94){(d)}
    \includegraphics[width=0.32\textwidth]{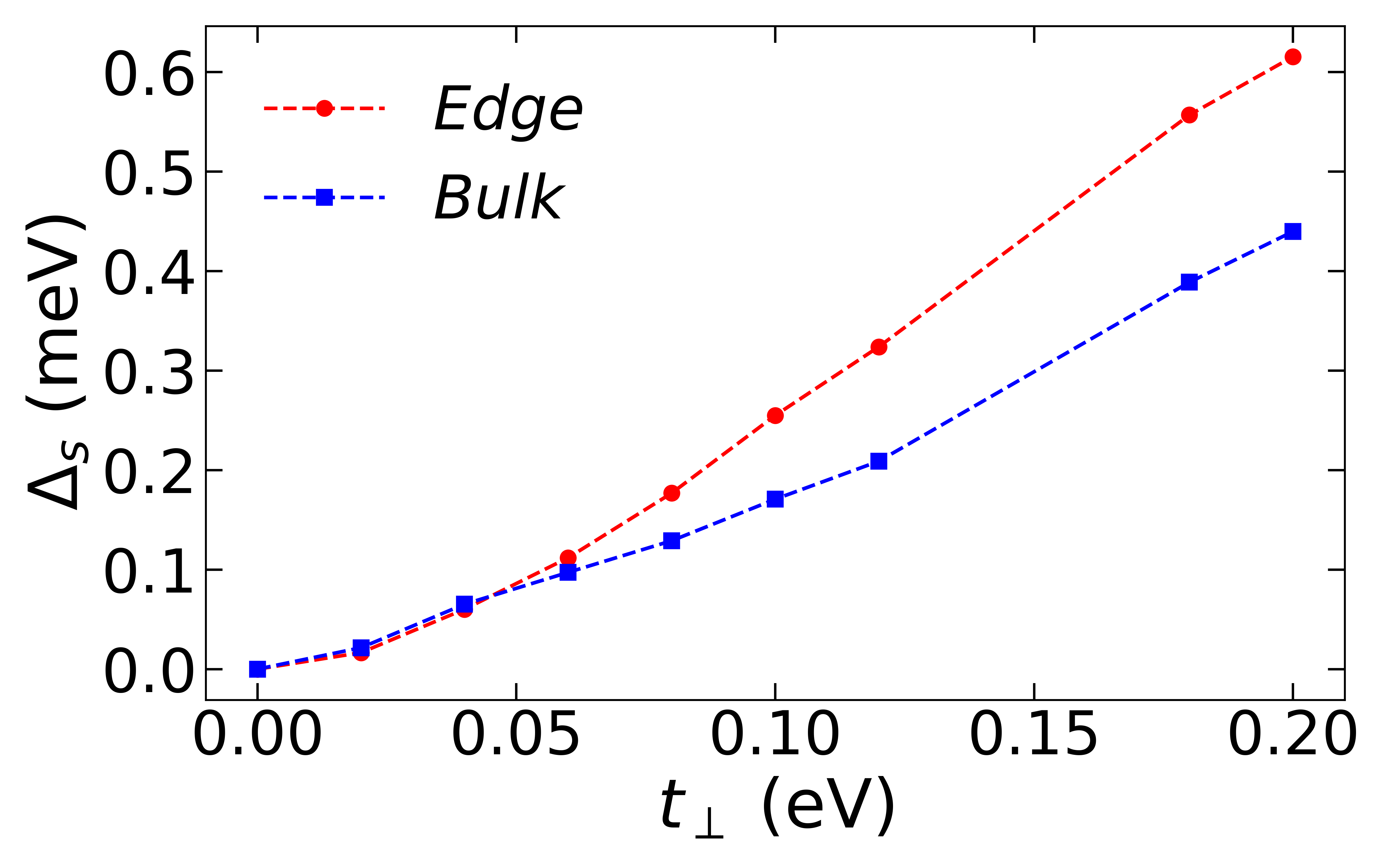}
    \put(-163,94){(e)}
    \includegraphics[width=0.32\textwidth]{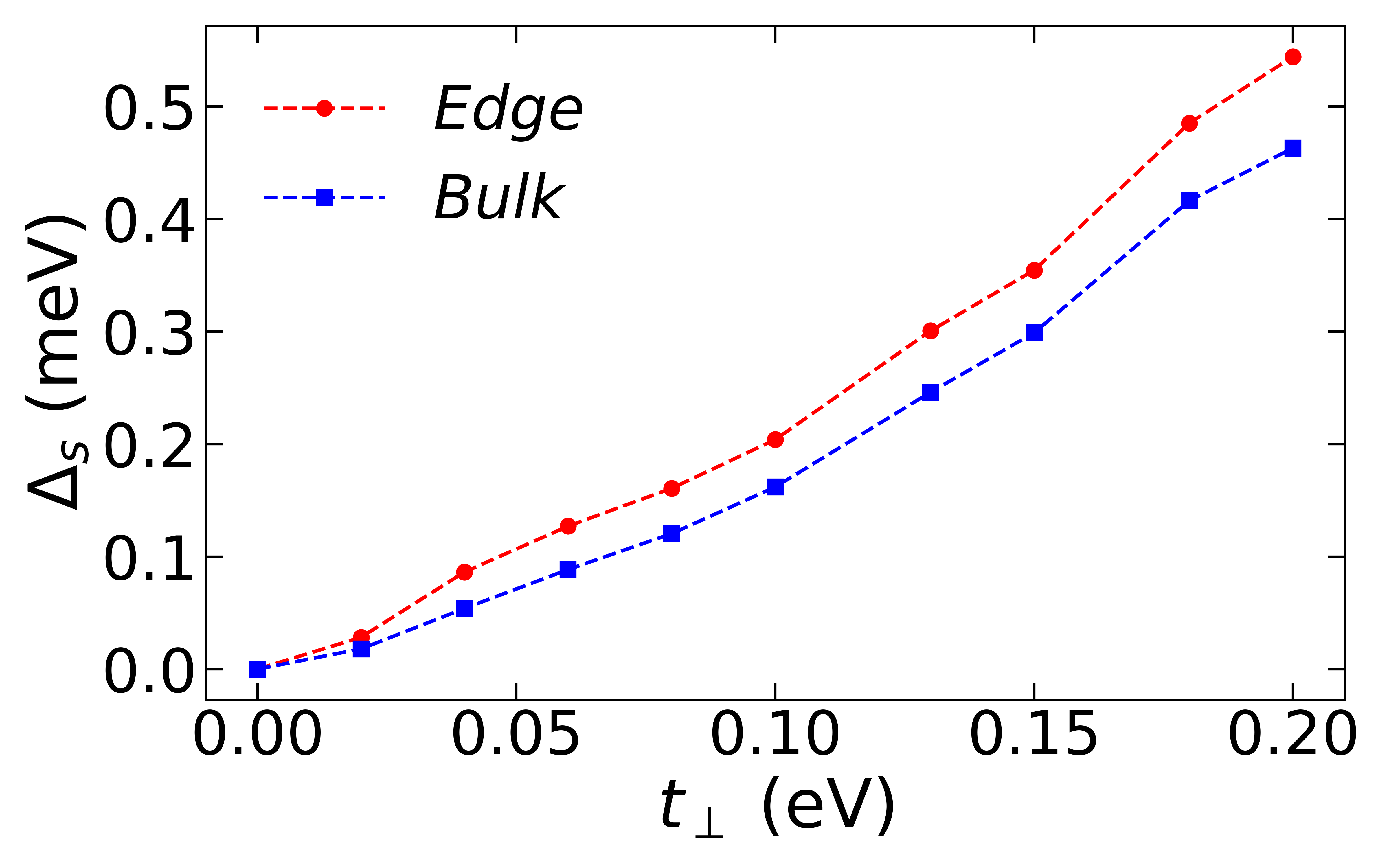}
    \put(-163,94){(f)}
   \caption {The spatial variation of induced s-wave superconducting gap in \wtt~for three different edges with interface coupling $t_\perp$ = 0.06 eV are shown in the top panel ((a) - (c)). Four different colors are indicating four ions (red = W, blue = W, green = Te, black = Te ) in a unit cell of \wtt. Variations of the induced gap (unit cell averaged) for the edge (red line) and the bulk (blue line) with the coupling strength ($t_\perp$) for different edges are shown in the bottom panel ((d) - (f)).}
    \label{singlet}
\end{figure*}
The van-der Waals heterostructure between 2H-NbSe$_2$ and 1T'-WTe$_2$ has been modeled with the nearest neighbor hopping $t_{\perp}$ that acts as a single parameter representing the interlayer coupling strength. It is assumed that this parameter would generally characterize the variation in interlayer coupling strength seen in this heterostructure under experimental conditions. The interlayer hopping includes the contributions from an effective hybridization between the Nb d$_{z^2}$ orbitals on \nbh~and the effective W [$d_{x^2 - y^2}$,$d_{z^2}$] orbitals in monoclinic \wtt~leading to a metallic edge state whose local density of state reasonably agrees with our experimental observations \cite{tao22}.

\section{Results}

In Fig. \ref{normal} we show the variation in the local density of states for both the insulating bulk (black) and metallic edge-state (red) at the edge (b) for interface coupling $t_\perp$ varying between 0.02 eV to 0.2 eV. For small interaction strength $t_{\perp}=0.02$ eV (See Fig.~\ref{normal} (a)), the spin Hall gap ($\sim$ 56 meV) is clearly visible for the bulk spectrum (black line), and the edge state LDOS shown in red line also retains features similar to the corresponding conducting edge state for pristine \wtt~(see Fig. \ref{pristine_bands}). Due to increasing hybridization at the interface, the spin Hall gap starts shrinking (see Fig. \ref{normal}(a) - \ref{normal}(f)) developing a V-shaped structure indicative of the formation of a significant momentum-dependent anisotropy in the formation of hybridization induced low energy electronic states. 

For larger interaction strengths $t_{\perp}>0.15$ eV the DOS at the Fermi energy becomes significantly larger. As shown in Fig. \ref{normal}, a finite DOS develops at the Fermi energy (E$_F$) even at $t_{\perp}=0.06$ eV. It can also be noted in Fig.~\ref{normal} that with increasing interaction strength, a sharp DOS peak develops in the valence band at around Energy, $\omega =-40$ meV, that is not present for weak interlayer hybridizations. This is likely due to a bulk contribution from the formation of a flat band at the $\Gamma$-point in the hybridized band structure around this energy, as shown in Fig.~\ref{bulk_bands}(b), where the deep blue lines are indicating the modified \wt~bands and the light grey lines are the \nb~bands. To gain insight into the changes in electronic structure arising from hybridization, we have included the pristine \wt~band structure in blue lines in Fig. \ref{bulk_bands}(a), overlaid with the pristine \nb~band structure in grey lines. These results from calculations of the normal state DOS motivate us to look at the corresponding effect of interlayer coupling on the superconducting state in \wt.

\begin{figure}[t]
    \centering
    \includegraphics[width = \columnwidth]{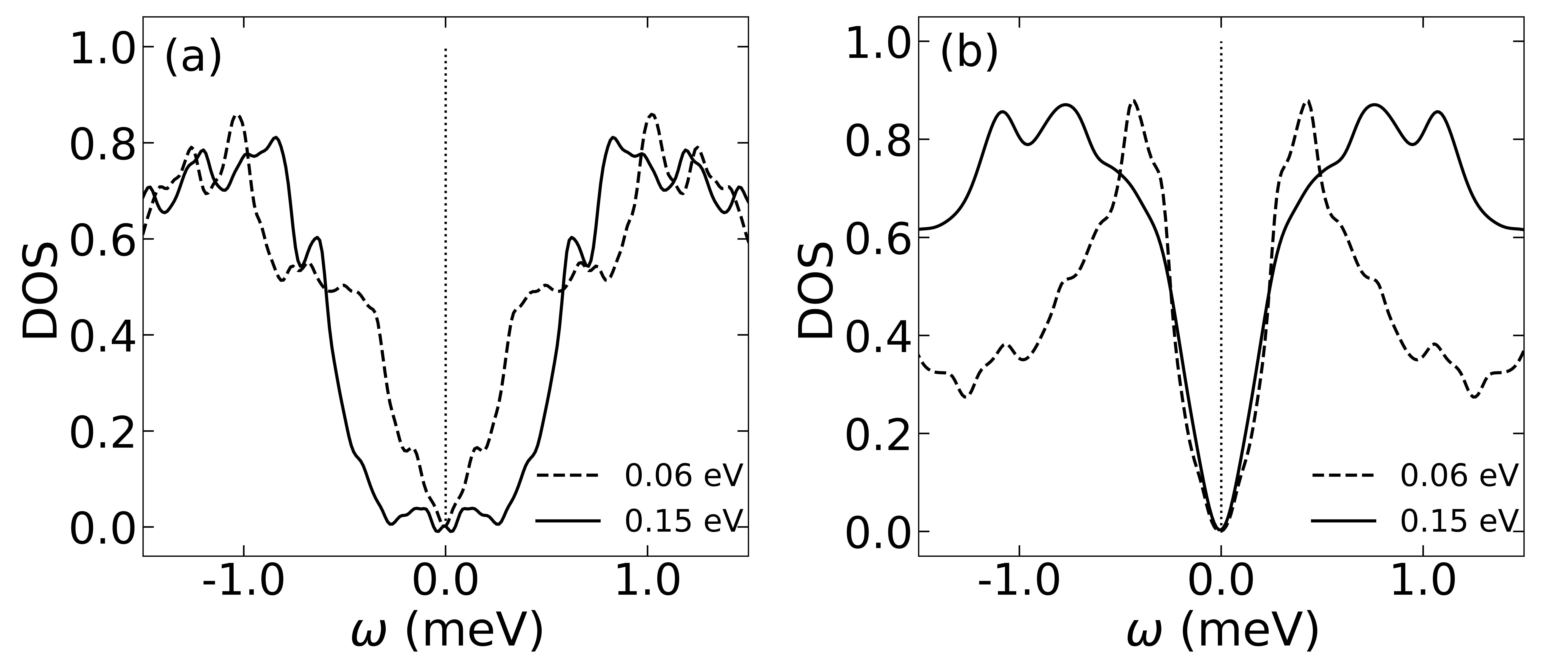}
    \caption{Comparison of the superconducting state DOS of the (a) X-edge (edge (a)) and (b) Y-edge (edge (b)) for the two interface coupling strengths 0.06 eV (dotted line) and 0.15 eV (solid line). }
    \label{edge_state_dos}
\end{figure}

 In Fig.~\ref{singlet} we show the spatial variation of the induced s-wave gap on \wt~with interlayer interaction strength of $t_\perp$ = 0.06 eV for the three different edge terminations. The plots present the superconducting gaps resolved over the 4 ions (2 W and 2 Te) in the unit cell. As shown in Figs.~\ref{singlet}(a), \ref{singlet}(b), and \ref{singlet}(c) that refer to the edges (a), (b), and (c) respectively, the induced gap over the W ions are larger than the gaps over the Te ions by approximately a factor of two.
 Additionally, we also find that the magnitude of the edge state superconducting gap is sensitive to the particular atomic termination, likely due to the variation in the edge state local density of states discussed in Fig.~\ref{edge_state_dos}. In particular, it is interesting to note that the largest induced superconducting gap is over the W ions for edge (a) that shows a linear edge state dispersion.

 In order to identify the variation of the induced superconductivity with interlayer hybridization, the plots in the lower panel (See Fig.~\ref{singlet}(d) - Fig.~\ref{singlet}(f)) are showing the variation of the induced superconducting order parameter (unit cell averaged) at the edge and the bulk for increasing hybridization strength from 0.02 eV to 0.2 eV. We find that as the interaction strength increases, as expected the edge state superconductivity, which is mainly dominated by the W ions, also increases for all of the terminations. Even at the edge, the superconducting gap of W ions remains larger than the Te ions with increasing $t_{\perp}$. We have also noticed that the difference between the bulk and edge gaps increases as the interaction strength increases. It depicts how the interaction strength is responsible for the larger edge state superconducting gap. Another feature that we have observed from the spatial variation plots is the small amplitude oscillations (See Figs. \ref{singlet} (b), \ref{singlet} (c)) present particularly for the edge (b), and edge (c). These oscillations are formed by interference effects due to the presence of the edge, and are in general suppressed with increasing strength of induced superconducting gap (or increasing interlayer hybridization). However, in general, these oscillations of the orbital resolved superconducting gap have significantly large coherence lengths in our heterostructure calculation.  

\begin{figure}
    \centering
    \includegraphics[width =\columnwidth]{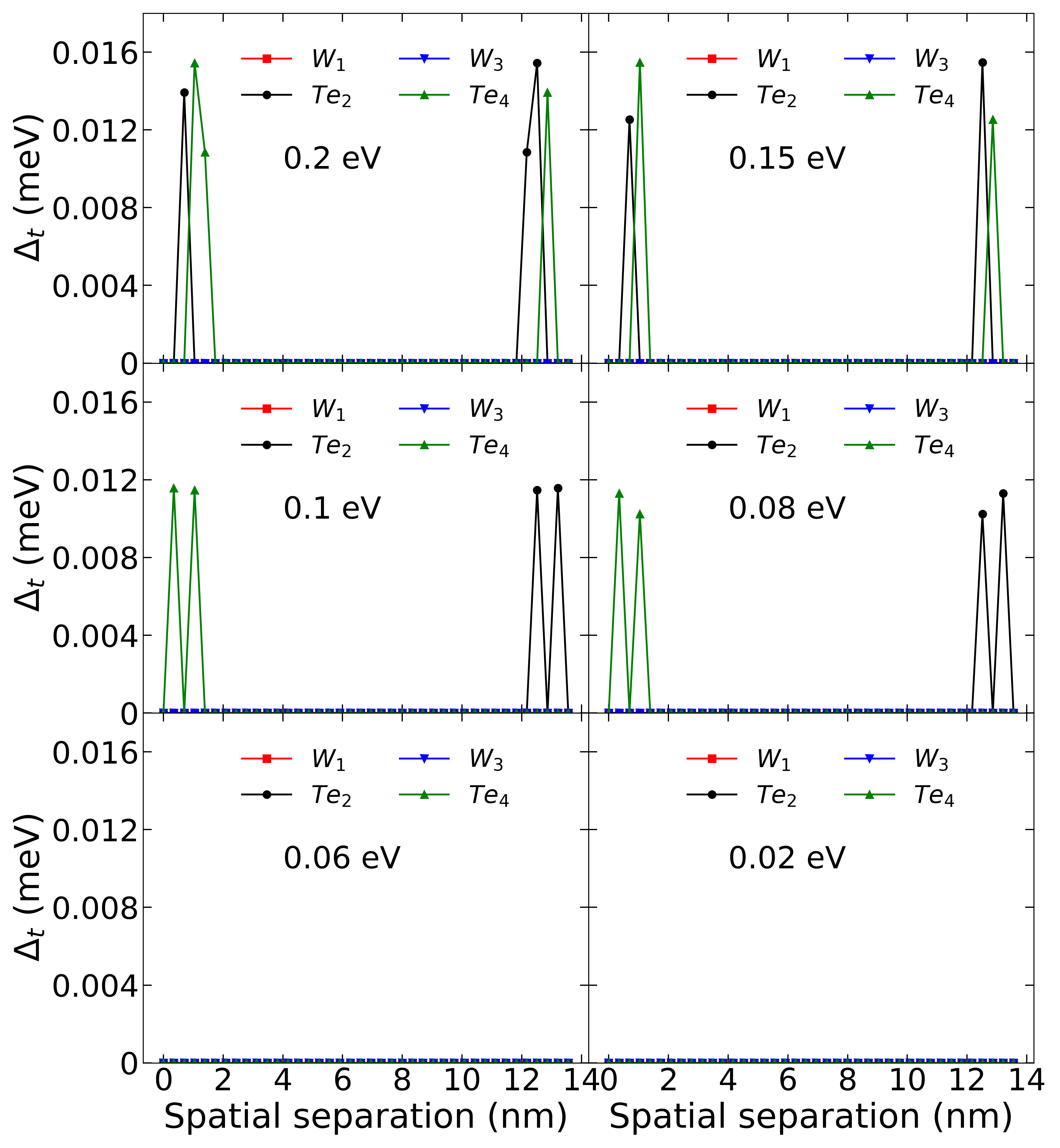}
    \caption{Spatial variation of Fu-Kane triplet for the Y-edge (edge - (b)) of \wtt~for different interface coupling as mentioned in each figure. Colors (red = W, blue = W, green = Te, black = Te) are indicating different atoms in a unit cell of \wtt. }
    \label{triplet_case_c}
\end{figure}

\begin{figure}
    \centering
    \includegraphics[width =\columnwidth]{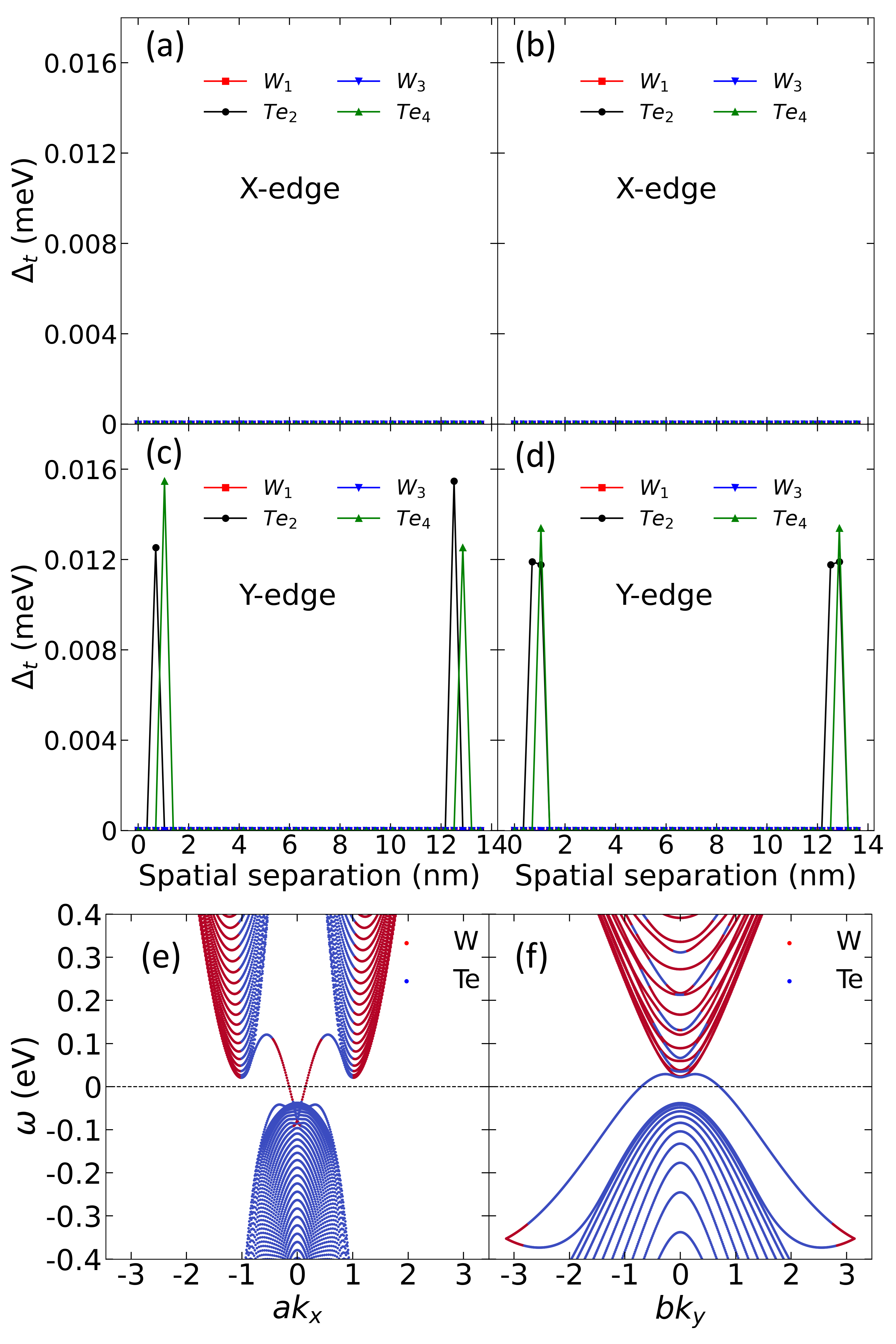}
    \caption{Plot (a-d) show topological triplet formation at the edge states. Note that although we have included results from an additional X-edge termination, topological triplet state is only obtained for a Y-edge. Panel (e)-(f) show the orbital resolved edge state band structure for an X-edge (e) and Y-edge (f). Irrespective of the termination, we find the conducting edge states band for an X-edge is dominated by W $d_{x^2-y^2}$ orbital and Y-edge is dominated by Te $p_x$ orbital contribution.} 
    \label{fig_9}
\end{figure}

 \begin{figure}[t]
    \centering
    \includegraphics[width =0.5\textwidth]{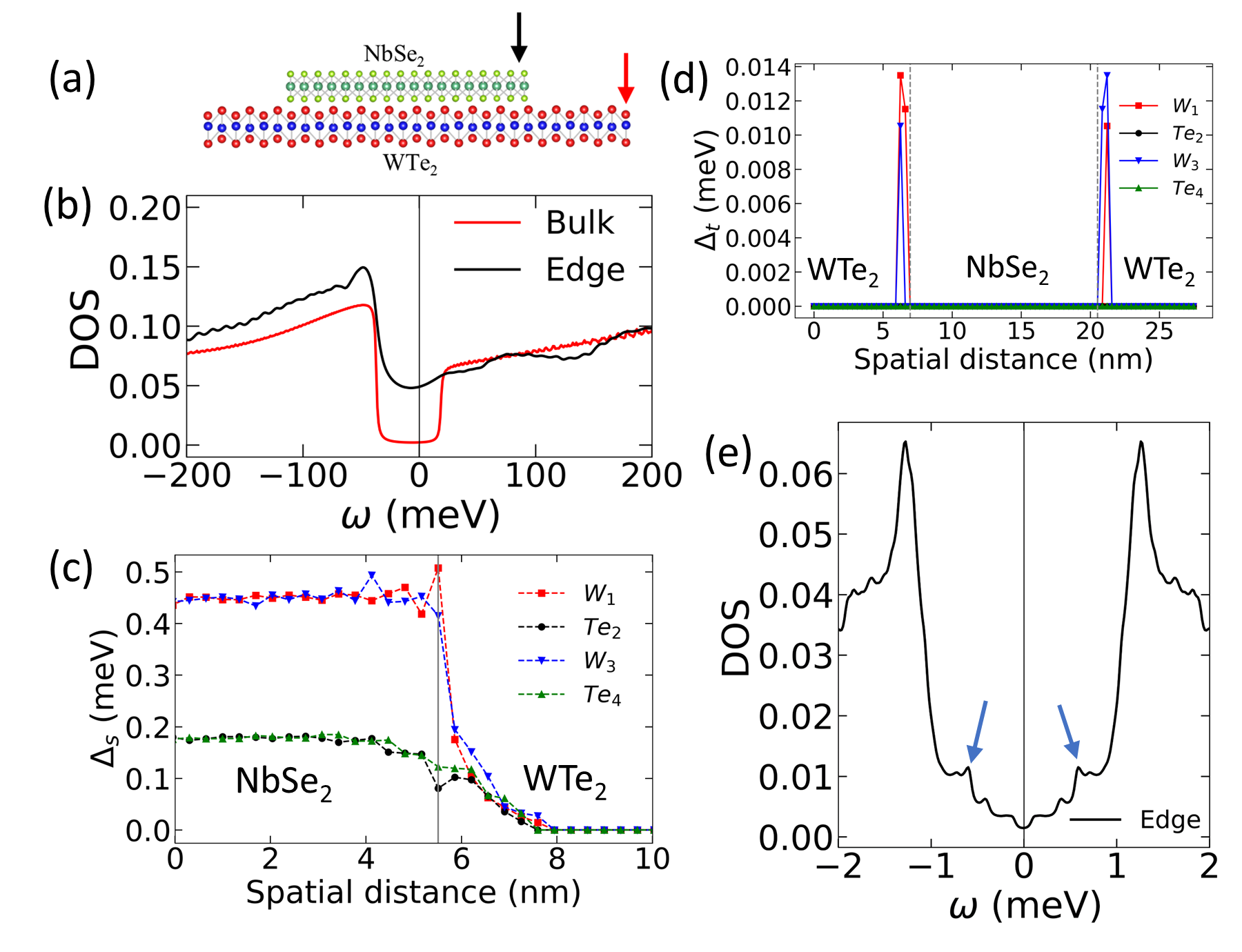}
    \caption{(a) Schematics of the heterostructure showing \nb~on top of \wt. (b) Calculated normal state DOS spectrum of \wt~for the edge (black line) and the bulk (red line) as indicated by the black and red arrows in Fig. (a) respectively. (c) Spatial variation of the orbital resolved induced s-wave superconductivity in \wt. The vertical gray line is indicating the edge of \nb~over \wt. The Left of this line is showing the induced gap on \wt~which is below \nb~and the right part is showing the gap on the rest of the pristine \wt. (d)} Spatial variation of the calculated Fu-Kane triplet state which is highly localised at the edges of the heterostructure. \textcolor{blue}{(e)} Calculated DOS for the superconducting state in \wtt~at one edge of the heterostructure. 
    \label{reverse}
\end{figure}
 
In Fig.~\ref{edge_state_dos} (a), and Fig.~\ref{edge_state_dos} (b) we present the calculated DOS at the edge of \wt~for different interlayer couplings 0.06 eV (dashed line) and 0.15 eV (solid line) for the X-edge(edge (a)) and Y-edge(edge (b)) respectively. The DOS has the typical features of a multi-gap induced superconducting state, and the particle-hole symmetric coherence peaks representing the small gaps agree well with the calculated mean field gaps on the W and Te ions at the edge states shown in Fig.~\ref{singlet}(a), and Fig.~\ref{singlet}(b). In particular, the edge (b) forms a V-shaped superconducting LDOS representing a significant superconducting gap anisotropy although the multi-gap features for this edge are weaker than for edge (a).
 
We further explore the non trivial nature of the electronic structure in the presence of hybridized bands by looking for the presence of an effective triplet superconducting gap in \wtt~following the formalism provided by Fu and Kane \cite{fu2008superconducting}. The presence of such an effective triplet order would be indicative of the presence of a chiral edge state electronic structure. In Fig.~\ref{triplet_case_c} we show the induced triplet order for various interlayer hybridization strengths. For weaker hybridizations, the induced triplet does not exist but becomes appreciable for stronger interlayer coupling. We also find that the triplet order is highly localized at the edge state, indicating that the chiral nature of the edge state electronic band survives in the presence of interlayer hybridization. As shown in Fig. \ref{fig_9}(a-d) we find that irrespective of the terminating atoms at the edge, the Te dominated topological triplet state is obtained in our self-consistent calculations for the Y-edge terminations but do not exist for the X-edge. Note that since the Te is much more weakly coupled to the underlying NbSe$_2$ compared to the W atoms, the Te dominated bands retain a significant amount of their helical character in the edge states. Therefore any contribution from the Te orbitals (or atom) at the edge state will be expected to show a dominant topological triplet character. As shown in Fig. \ref{fig_9}(e) and (f), the anisotropy in topological triplet between X-edge and Y-edge can be understood as arising from the dominant orbital contribution of the Te $p_x$ orbital on Y-edge band whereas the X-edge band is dominated by W orbital contribution. We find that this anisotropy in the orbital content of the band structure is independent of the atomic edge termination and is primarily governed by the directional anisotropy between X and Y-terminations.

In addition to the possibility of tuning the interlayer hybridization with spacer layers, the heterostructures with stronger interlayer hybridization could be used to generate cleaner interfaces. In Fig.~\ref{reverse} we show an example of an inverted heterostructure geometry where a superconducting island of \nb~is placed over a \wt~monolayer. In Fig.~\ref{reverse}(b) we can see that in this case although the conducting edge state has a significant DOS, the \wt~regions further from the island retain their QSHI gap. In fact, this interface can be expected to show non-trivial edge state properties (where the edge is defined by the termination of \nb) since it separates a topologically non-trivial regime showing the QSHI gap from a topologically distinct regime where the closed QSHI gap is induced by a strong hybridization between \nb~and \wt. In Fig.~\ref{reverse}(c), we show that the self-consistent calculation results considering an intrinsic superconducting state on \nb~leads to a sizable superconducting gap of $\Delta \sim 0.6$meV at the \wt~edge state with the interface coupling strength $t_\perp = 0.15$ eV. This gap is clearly visible in the DOS calculations shown in Fig~\ref{reverse}(e) at the \wt~edge. Additionally, the spatial variation of the calculated Fu-Kane type triplet gap has been shown in Fig~\ref{reverse}(d). We find that the triplet is highly localised at the edges of the heterostructure indicating that the chiral behaviour of the edge state electronic bands survives for this heterostructure arrangements.

\section{Conclusion}
Based on experimentally observed variations in interlayer coupling in \wtt/\nbh~heterostructure, we have reported a theoretical study of the evolution of electronic structure with interlayer coupling in the normal and superconducting state. The interlayer hybridization suppresses the quantum spin Hall gap of \wtt~in the normal state. The electronic structure of the edge state of \wt~shows significant hybridization with the Nb d$_{z^2}$ orbitals even for relatively weak interlayer coupling of $t_{\perp}=0.06$ eV although the density of states of the edge state remains larger than the corresponding bulk values which are consistent with recent experimental findings \cite{tao22}.

For the superconducting state, we find that even in the presence of hybridization the induced superconducting gap at the edge remains stronger than the gap induced on the surface. We also find that a weaker interface hybridization leading to a hopping strength of around 0.05-0.07 eV might be a reasonable regime to get a suppressed bulk DOS and a reasonable gap induction at the edge. However, the nature of the induced gap at the edge state depends on the particular atomic termination or edge orientation. Additionally, we also identify the topological features of the superconducting state by calculating the effective Fu-Kane type chiral spin-triplet states and showed that they are highly localized at the edge, indicating the helical nature of the edge state survives in the presence of interlayer coupling. The induced s-wave superconducting gap is dominant on the W orbital compared to the Te due to the larger hybridization of the W orbital/atom with the underlying \nb~substrate. The stronger hybridization leads to a larger orbital mixing between the \nb~and W bands and consequently leads to a stronger induced superconducting gap on the W atoms. Conversely, as discussed above the weaker hybridization with the Te orbitals helps retain the helical nature of the Te bands at the edge states and supports topological triplet superconducting gap on the Te atoms at the Y-edge. We find the qualitative behavior of the induced even parity and topological odd parity superconducting state is robust against small changes in the electronic doping levels.

Finally, we propose that for certain heterostructure geometries, the strong interlayer hybridization between a superconductor and 2D topological insulator can be beneficial for generating nontrivial topological edge state properties. In particular, we discuss this scenario for a \nb~island placed over a \wt~monolayer. The monolayer could be fabricated over a graphitic substrate that is known to be weakly coupled to the monolayer and allows the QSHI state on \wt~to remain intact. A similar scenario may be applicable to other 2D topological materials with conducting edge states or 3D topological superconductors with conducting surface states that have a strong inter-layer hybridization to a superconducting material.

To sum up, in this study, we have investigated the nature of induced superconductivity in a realistic model, emphasizing the significant role of orbitals, edge terminations, and termination directions in identifying signatures of topological superconductivity. Our findings highlight the importance of tunneling into specific orbitals in STM experiments, particularly those associated with the Te atoms in the \wtt/\nbh~heterostructure. Moreover, our results reveal anisotropy in the topological superconductivity between X- and Y-edge terminations for the \wt~and \nb~heterostructure. Overall, our findings indicate that the existence and anisotropy of topological superconductivity in the heterostructure between \wtt and \nbh are inherently influenced by the underlying geometry dependence of both bulk and edge states. This phenomenon of geometry-dependent topological superconductivity has also been explored in other topological materials \cite{chou2021geometry, salamone2022curvature, trang2020conversion}.

These results suggest that for the heterostructure system studied in this work, the directional anisotropy in the topological features can be probed in tunneling as well as transport experiments. For example, near X and Y-edges where topological triplet state only forms at the latter edge, STM measurements probing magnetic vortices can be expected to find significant anisotropy in the in-gap states. The observation of topological superconductivity only along one edge direction could make it interesting to probe corner states connecting perpandicular edges in monolayer \wtt. Such hybridized heterostructure systems can also be amenable to tunability of topological properties either by introduction of spacer layers or through formation of an alternate geometry, an example of which has been discussed in this work.

\appendix\section{Self-consistent BdG equations
with superconductivity}
The Hamiltonian for the heterostructure in real space is,
\begin{eqnarray}
	\mathcal{H}&=&\mathcal{H}^{0}_{N}+\mathcal{H}^{0}_{W}+\mathcal{H}_{W-N} +\mathcal{H}^{N}_{SC}
\end{eqnarray} 
Where the four different parts are given,
\begin{eqnarray}
	\mathcal{H}^{0}_{N}&=&\sum_{iljl'\sigma}t^{ll'}_{ij}c^{\dag}_{il\sigma}c_{jl'\sigma} + h.c \nonumber\\
	\mathcal{H}_{W-N}&=&t_{\perp} \sum_{ij\sigma} c^{\dag}_{i2\sigma}d_{j1\sigma} + h.c \nonumber\\
	\mathcal{H}^{0}_W&=&\sum_{\mu\nu ij\sigma\sigma'}t^{ij}_{\mu\nu\sigma\sigma'}d^{\dag}_{i\mu\sigma}d_{j\nu \sigma'} + h.c \nonumber\\
	\mathcal{H}^{N}_{SC}&=&  \sum_{il}\Delta^{N}_{il} c^{\dag}_{il\uparrow}c^{\dag}_{il\downarrow}+ h.c\nonumber
\end{eqnarray}

Here $\sigma$ is the spin index, $i,j$ are the site indices. $l$ is the orbital index for \nbh~and $\mu,\nu$ are the orbital indices for \wtt. $c_{il\sigma}$ is the electron annihilation operator for \nbh~and  $d_{j\nu \sigma}$ is the same for \wtt. $t_{ij}$ and $t_\perp$ are the intra-plane and inter-plane hopping parameters.
To make the calculation easier we Fourier transform the above Hamiltonian in one direction and keep the other direction in real space.

We define a parameter $\Tilde{h}_{ki\sigma j\sigma'}$, which contains all the hopping parameters ($t_{ij}$ and $t_{\perp}$) of the heterostructure in the above basis. The induced superconducting gap in the spin-singlet channel is evaluated self consistently by calculating the anomalous averages $\Delta^{W}_{i\mu}\sim \langle d^{\dag}_{i\mu \uparrow}d^{\dag}_{i\mu \downarrow}\rangle $, where $\mu=(1,2,3,4)$ represents the 2 Te, and 2 W atoms in the unit cell and $i$ is the unit cell index. Similarly the singlet superconducting gap in \nbh is calculated by $\Delta^{N}_{i\l}\sim V\langle c^{\dag}_{i\l \uparrow}c^{\dag}_{i\l \downarrow}\rangle $. $V$ is the pairing potential.

For an edge along a particular direction of \wt, the Bogoliubov-de Gennes (BdG) transformations involve quasi-particle operators,
	\begin{eqnarray}
	d_{ik\mu\sigma}&=&\sum_{n}(u^n_{ik\mu\sigma}\gamma_{n\sigma}-\sigma v^{n\star}_{ik\mu\sigma}\gamma^\dag_{n\sigma})\\
	d^\dag_{ik\mu\sigma}&=&\sum_{n}(u^{n\star}_{ik\mu\sigma}\gamma_{n\sigma}^\dag - \sigma v^n_{ik\mu\sigma}\gamma_{n\sigma})
	\end{eqnarray}

Here, $\mu$ is the orbital index, and n represents the quasi-particle index for the BdG Hamiltonian. Similar, expressions for the electron operator have been utilized for each layer in \nbh.

The full Hamiltonian in the matrix in the real space can be written with the superconducting blocks included in it,
\begin{eqnarray}
\mathcal{H}_{ij} = \begin{pmatrix}
\Tilde{h}_{i\uparrow j\uparrow} & \Tilde{h}_{i\uparrow j\downarrow} & 0 & \Delta_{ij}\\ 
\Tilde{h}_{i\downarrow j\uparrow}& \Tilde{h}_{i\downarrow j\downarrow}  & \Delta_{ji} &0\\
0 & \Delta^*_{ij} &-\Tilde{h}^*_{i\uparrow j\uparrow} &\Tilde{h}^*_{i\uparrow j\downarrow}\\
\Delta^*_{ji} & 0 &\Tilde{h}^*_{i\uparrow j\downarrow} & -\Tilde{h}^*_{i\downarrow j\downarrow}
\end{pmatrix}    
\end{eqnarray}

Here $\Tilde{h}_{i\sigma j\sigma'}$ contains the normal state part of the full Hamiltonian and $\Delta_{ij}$ is the superconducting part of the Hamiltonian.

We Fourier transform the above Hamiltonian in momentum space along one direction and keep the other direction in real space. Diagonalizing the above BdG Hamiltonian, we self-consistently calculate the electron density and superconducting gap at each lattice site for the multi-orbital Hamiltonian. The mean fields would in general be given by,

	\begin{eqnarray}
	n_{i\mu\sigma} = \frac{1}{N_{k}}\sum_{k,n}|u^n_{ik\mu\sigma}|^2 f(E_n)
	\end{eqnarray} 
	Here, $f(E_n)$ is the Fermi function.
As discussed above, the superconducting gap on NbSe$_2$ has been introduced with a pairing interaction term V. The self-consistent procedure leads to an induced even parity ($\Delta^{s}_{ij\mu}$) and odd parity ($\Delta^{t}_{ij\mu}$) order parameters on 1T$^\prime$-WTe$_2$. The induced superconducting gaps are obtained from the self-consistent solutions by calculating the following anomalous averages,
	
	\begin{eqnarray}
	\Delta^{s}_{ij\mu}=\frac{1}{N_{k}}\sum_{n,k=0}^{2\pi}[u^n_{ik\mu\uparrow}v^{n\star}_{jk\mu\downarrow}+u^n_{jk\mu\downarrow}v^{n\star}_{ik\mu\uparrow}]f(E_n)\nonumber\\
	\\
	\Delta^{t}_{ij\mu}=\frac{1}{2N_{k}}\sum_{n,k=0}^{\pi}[u^n_{ik\mu\uparrow}v^{n\star}_{jk\mu\downarrow}-u^n_{ik\mu\downarrow}v^{n\star}_{jk\mu\uparrow}]f(E_n)\nonumber\\
	\end{eqnarray}
Here, $N_{k}$ is the number of $k$ divisions which is typically taken to be 30000 points to achieve high resolution for the small gaps observed in the system.

In terms of expectation value of the electron creation ($c^\dag$)  and annihilation ($c$) operators all the above-mentioned mean-field quantities that represent the site and orbital resolved electron densities, and induced superconducting gap on \wt can be expressed as,
\begin{align}
    n_{i\mu\sigma}&= \frac{1}{N_k}\sum_k \langle d^{\dag\mu}_{i,k,\sigma}d^\mu_{i,k,\sigma}\rangle\\
    \Delta^s_{ij\mu} &=\frac{1}{N_k}\sum_k \langle d^{\dag\mu}_{i,k,\uparrow}d^{\dag\mu}_{j,-k,\downarrow}-d^{\dag\mu}_{i,k,\downarrow}d^{\dag\mu}_{j,-k,\uparrow}\rangle\\
    \Delta^t_{ij\mu} &=\frac{1}{N_k}\sum_k \langle d^{\dag\mu}_{i,k,\uparrow}d^{\dag\mu}_{j,-k,\downarrow}+d^{\dag\mu}_{i,k,\downarrow}d^{\dag\mu}_{j,-k,\uparrow}\rangle
\end{align}
 Where ‘i’ , ‘j’ are the real-space points along a particular direction that represent the unit cell index,  is the orbital index that run over the 8 orbitals, $\sigma$ is the spin index, and $N_k$ are the number of reciprocal lattice k points (which can be either $k_x$ or $k_y$ depending on the chosen lattice termination) perpendicular to the real space direction. Note that i equal to j here would correspond to the induced onsite s-wave superconducting gap.
 
\bibliographystyle{apsrev4-2}
\bibliography{ref}
\end{document}